\documentclass[11pt]{article}
\usepackage[utf8]{inputenc}
\usepackage{mathtools,amsmath,amssymb}
\usepackage{graphicx}
\usepackage{lmodern}
\usepackage{}
\usepackage[T1]{fontenc}
\usepackage{bbm}
\usepackage{bm}
\usepackage{appendix}
\usepackage{comment}
\usepackage{mathtools}
\usepackage{array}
\usepackage{amsthm}
\usepackage[version=4]{mhchem}
\usepackage{subcaption}
\usepackage{authblk}
\usepackage{accents}
\usepackage{enumerate}
\usepackage{todonotes}
\usepackage[colorlinks=true, allcolors=blue]{hyperref}
\usepackage[left=2cm,right=2cm,top=2cm,bottom=2cm]{geometry}
\usepackage{caption}
\numberwithin{equation}{section}

\newtheorem{definition}{Definition}

\theoremstyle{definition}

\usepackage{xspace}

\usepackage{graphicx}

\usepackage{subcaption} 

\allowdisplaybreaks
\usepackage{tikz}

\title{Uphill transport in competitive drift-diffusion models\\ with volume exclusion}

\author[1]{Francesco Casini\thanks{francesco.casini@kuleuven.be}}
\author[2]{Cristian Giardin\`a\thanks{cristian.giardina@unimore.it}}
\author[3]{Jacopo Nicolini\thanks{jacopo.nicolini@unimore.it}}
\author[3]{Luca Selmi\thanks{luca.selmi@unimore.it}}
\author[2]{Cecilia Vernia\thanks{cecilia.vernia@unimore.it}}
\affil[1]{Department of Physics and Astronomy, KU Leuven, Leuven, Belgium.}
\affil[2]{Department of Physics, Informatics and Mathematics, University of Modena e Reggio Emilia, Modena, Italy.}
\affil[3]{DIEF, University of Modena e Reggio Emilia, Modena, Italy.}
\date{\today}
\begin{document}
\maketitle
\begin{abstract}
This paper addresses uphill transport (defined as a regime in which particle flow is opposite to the prescriptions of Fick's diffusion) in drift-diffusion particle transport constrained by volume exclusion. Firstly, we show that the stationary hydrodynamic limit of a multispecies, weakly asymmetric exclusion process (SHDL) naturally predicts precisely characterized uphill regimes in the space of external drivings.

Then, with specific reference to systems of oppositely charged particles, we identify well-defined model hypotheses and extensions whereby the SHDL converges to the modified Poisson-Nernst-Planck model, thus bridging the gap between exclusion-based particle models and continuum descriptions commonly used in engineering. The merits and limitations of the models in describing the particle fluxes and predicting uphill transport conditions are investigated in detail with respect to the adopted approximations and simplifications. 

The results demonstrate the persistence of uphill transport phenomena across modeling scales, clarify the conditions under which they occur, and suggest that uphill transport may play a significant role in nanoscale electrolytes, confined ionic and iontronic devices, and membrane-based technologies.
\end{abstract}
\vspace*{2truecm}
\newpage
\section{Introduction}
Drift-diffusion processes are fundamental to the description of mass and charge transport in a wide range of physical, chemical, and biological systems. For charged entities, e.g., ion migration in electrolytes and biological membranes and carrier motion in semiconductor devices, these transport phenomena are typically modeled by \textit{Poisson–Drift-Diffusion} equations, equivalently denoted \textit{Poisson–Nernst–Planck} (PNP) equations,
which describe the self-consistent electrostatic interaction between charged particles, concentration gradients, and electric fields. 

Common formulations of these models often neglect exclusion effects originating, e.g., from 
the finite volume occupied by solvated ions in electrolytes. 
In systems characterized by high concentrations, nanoscale confinement, or strong electrostatic coupling, this approximation can break down, leading to nontrivial transport properties that depart from pure drift-diffusion. 
At present, there is growing interest in studying uphill transport - that is, fluxes directed in accordance with the sign of the gradient of concentration - for multi-component systems, both from a phenomenological perspective \cite{krishna2015uphill,krishna2017highlighting,krishna2019diffusing} and in the mathematical-physics literature \cite{brzank2006boundary,colangeli2017microscopic,floreani2022switching,casini2023uphill,Colangeli_2023}. 

In this work, we investigate the emergence of \textit{uphill transport} in systems governed by drift-diffusion dynamics in the presence of volume exclusion. In Section\,\ref{Sect:UphillPDEs}, starting with the microscopic exclusion model introduced in previous studies \cite{casini2025large}, we analyze its hydrodynamic limit (HDL), which provides a set of coupled nonlinear partial differential equations for two species subject to drift, diffusion, and mutual exclusion in the absence of electrostatic interaction. 
The stationary (time-independent) solution of this system (hereafter denoted SHDL) with fixed boundary concentrations yields a non-equilibrium steady state with non-zero particle fluxes. 
By systematically exploring the parameter space of external fields and boundary concentrations, we identify conditions under which partial and global uphill transport regimes emerge, driven by the competition between boundary-induced diffusion and field-induced drift.

To link this theoretical framework with real physical systems, in Section\,\ref{Sect:Drift-diffusion models for charged particles} we establish a correspondence between the SHDL model and the \textit{modified Poisson–Nernst–Planck} equations (mPNP), which describe the electrostatic coupling and transport of charge entities with exclusion effects under general non-equilibrium conditions. 
 
In Section\,\ref{Section-comparison-OutOfEq}, through analytical and numerical comparisons, we demonstrate the mutual consistency of the SHDL (possibly endowed with the Poisson equation) and the mPNP models. Then, we identify the regions of the parameter space where these models predict uphill transport behaviors.

Finally, in Section\,\ref{Sect:Uphill-Section} we apply these models to a prototypical case study of charged particle transport: an ion-selective membrane between two electrolytes.

For this system, we examine how boundary conditions, volume exclusion (also called steric) effects, and electrostatic coupling can induce or suppress uphill transport, revealing parameter ranges where this behavior, opposite to Fick's law, is expected to prevail. 

The results, summarized in the conclusive Section\,\ref{Sect:Conclusions}, highlight the generality of uphill transport as a manifestation of competitive drift-diffusion mechanisms in systems of electrostatically interacting charged entities with volume exclusion and emphasize its potential relevance in nanostructured and electrochemical devices.

\section{Uphill transport in exclusion models with drift}
\label{Sect:UphillPDEs}

\subsection{Multispecies Weakly Asymmetric Simple Exclusion process}\label{subsect:microscopic-model}
\begin{figure}[hb]
    \centering
    \includegraphics[width=0.5\linewidth]{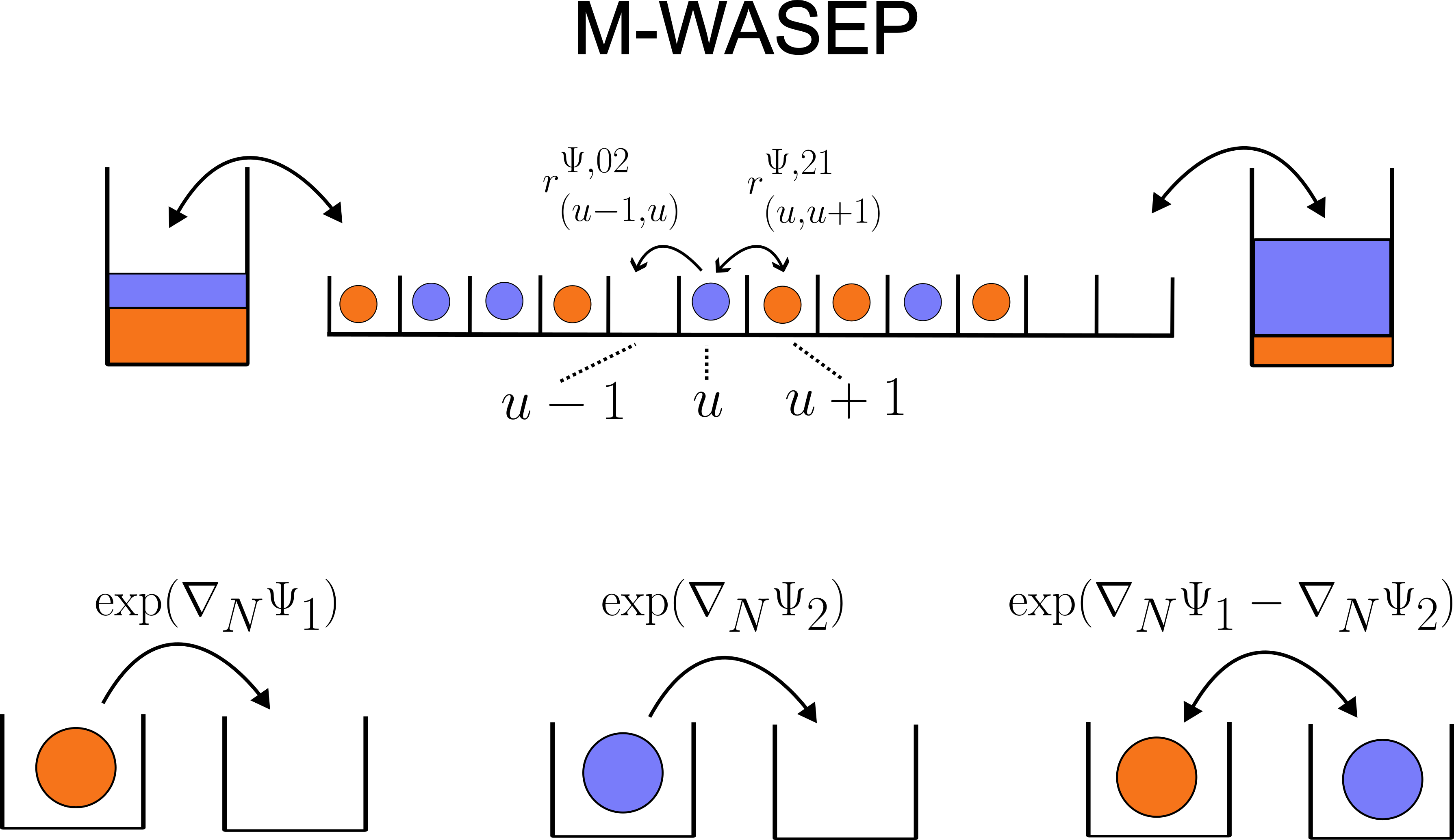}
    \caption{Representation of the microscopic Markov model with two particle species (blue and red) hopping across sites with weakly asymmetric probability rates.}
    \label{fig:sketch_wasep}
\end{figure}

To introduce the topic of our investigation, we recall the 
multispecies weakly asymmetric 
simple exclusion process (M-WASEP),
introduced in \cite{casini2025large} in the context of large deviations
of the corresponding symmetric  model.
This is a microscopic model
where particles jump on a lattice of finite length $\lfloor N\xi_u \rfloor$, with $\xi_u>0$ the macroscopic length of the system and $N$ the number of particles. Here (and throughout the paper) we consider the case of two species in a boundary-driven setup, with injection and removal of both particle types at the boundaries.
Figure \ref{fig:sketch_wasep} represents the model in one-dimension.

We consider the stirring version
of the model, in which particles
of different types can swap.
We denote by $\eta^{u}=(\eta_{0}^{u},\eta_{1}^{u},\eta_{2}^{u})$ the occupation vector at site $u$. 
Here, $\eta_{\alpha}^{u}\in \{0,1\}$ represents the number of particles of species $\alpha$ located at site $u$. We require the so-called exclusion constraint, which allows at most one particle per site, i.e., for each site $u
$
we assume that
\begin{equation}\label{exclusionConstraint}
    \eta_{0}^{u}+\eta_{1}^{u}+\eta_{2}^{u}=1\,.
\end{equation}
This constraint implies that the species with the label $0$ plays the role of vacancies. In other words, the configuration of the process only depends on the species with labels $1$ and $2$.
We denote by $\bm{\eta}=(\eta^{1},\ldots,\eta^{N})$ the occupation variable for the whole chain and by $(\bm{\eta}(\tau))_{\tau\geq 0}$ the Markov process.

The continuous time process evolves its configurations according to transitions between nearest neighbors, for which we associate transition rates. Let $\bm{\Psi}(\xi,\tau)=(\Psi_{1}(\xi,\tau),\Psi_{2}(\xi,\tau))$ be two smooth functions of space and time, with $\xi$ the continuous space variable. 

Then, for $i\in \{1,2\}$ and with rate
\begin{equation}
    r^{{\bf\Psi},i0}_{(u,u+1)}=\exp\Big(\nabla_N\Psi_{i}(\tfrac{u}{N},\tau)\Big)\eta_{i}^{u}(1-\eta_{1}^{u+1}-\eta_{2}^{u+1})
    \label{eqn:rate}
\end{equation}
a particle of type $i$ is moved from $u$ to $u+1$. With rate 
\begin{equation}
    r^{{\bf\Psi},0i}_{(u,u+1)}=\exp\Big(-\nabla_N\Psi_{i}(\tfrac{u}{N},\tau)\Big)(1-\eta_{1}^{u}-\eta_{2}^{u})\eta_{i}^{u+1}
\end{equation}
a particle of type $i$ is moved from $u+1$ to $u$. Finally, for $i,j\in \{1,2\}$ and with rate 
 \begin{equation}
    r^{{\bf\Psi},ij}_{(u,u+1)}=\exp\Big(\nabla_N(\Psi_{i}(\tfrac{u}{N},\tau)-\Psi_{j}(\tfrac{u}{N},t))\Big)\eta_{i}^{u}\eta_{j}^{u+1}
\end{equation}
a particle of species $i$ at site $u$ and a particle of species $j$ at site $u+1$ are swapped. Above, we denoted by $\nabla_N$ the discrete gradient, i.e., 
\begin{equation}
    \nabla_N\Psi_{i}(\tfrac{u}{N},\tau) = \Psi_{i}(\tfrac{u+1}{N},\tau) - \Psi_{i}(\tfrac{u}{N},\tau)\,.
\end{equation}
Informally, these transition rates describe the transition probability for the system to go from a configuration to another one in an infinitesimal time interval $d\tau$. We observe that if $\Psi$ is constant in space, then the transition rates reduce to the symmetric multispecies stirring process \cite{casini2024duality,vanicat2017exact}. Therefore, $\nabla_{N}\Psi_{i}$ plays the role of asymmetry since it produces a bias in the jump rates.  We say that the process is weakly asymmetric because, for $N$ large enough, the asymmetric part of the transition rates is proportional to $1/N$. 
We also remark that in the multispecies context, the model with stirring and that without stirring can have substantially different properties, see e.g.
\cite{quastel1992diffusion} for the multispecies symmetric exclusion process without stirring.

\subsection{Hydrodynamic limit} 

\label{SubSect:PDEs} 
In \cite{casini2025large}, the authors proved the hydrodynamic limit of the weakly asymmetric multispecies stirring process described in Section \ref{subsect:microscopic-model}. 
Intuitively, this consists of showing that the empirical densities of the microscopic occupation variables $\eta_{1}^{u},\eta_{2}^{u}$ converge (under diffusive space-time  scaling $\frac{u}{N} \to \xi, \frac{\tau}{N^2}\to t$ in the limit $N\to\infty$) to functions $\rho_{1}(\xi,t),\rho_{2}(\xi,t)$ satisfying the so-called hydrodynamic equations\footnote{One has to specify the initial conditions and the boundary conditions. For the initial condition, one typically assumes the so-called local equilibrium property at the initial time. Then, proving the hydrodynamic limit amounts to showing that local equilibrium is preserved over time, with the density of the evolving measure governed by the solution of the limiting equations (for details see \cite{demasiPresutti,kipnisLandim}). For the boundary condition we refer to the discussion below.}.
These limiting partial differential equations, hereafter denoted hydrodynamic limit model (HDL), read \cite{casini2025large}: 
\begin{equation}\label{eqn:HD-equations LDP_grouped}
\begin{split}
	\partial_{t}\rho_{1}=&\partial_{\xi\xi}\rho_{1}-2\partial_{\xi} \left(\rho_{1}(1-\rho_{1})\partial_{\xi} \Psi_1\right)+2\partial_{\xi} \left(\rho_{1}\rho_{2}\partial_{\xi} \Psi_2\right)
    \ , \\
	\partial_{t}\rho_{2}=&\partial_{\xi\xi}\rho_{2}-2\partial_{\xi} \left(\rho_{2}(1-\rho_{2})\partial_{\xi} \Psi_2\right)+2\partial_{\xi} \left(\rho_{1}\rho_{2}\partial_{\xi} \Psi_1\right)
    \ ,
    \end{split}
\end{equation}
where $\rho_1(\xi,t)$ and $\rho_2(\xi,t)$ are normalized volumetric densities, i.e., normalized concentrations. The unknown functions $0 \leq \rho_{1}\leq 1$ and $0 \leq \rho_{2}\leq 1$ have to satisfy the additional constraint (inherited from the exclusion constraint \eqref{exclusionConstraint})
\begin{align}\label{exclusion-macro}
	0\leq \rho_{1}(\xi,t)+\rho_{2}(\xi,t)\leq 1\qquad\xi\in [0,\xi^{R}],\,\;\,  t\geq 0\,.
\end{align}

The functions $\Psi_{1},\Psi_{2}: [0,\xi^{R}]\times \mathbb{R}_{+} \to \mathbb{R}$ (introduced for the microscopic model in Section \ref{subsect:microscopic-model}) are viewed as potentials of two external fields given by $-\partial_{\xi} \Psi_{1}$ and $-\partial_{\xi} \Psi_{2}$, respectively. In the most general case, these potentials are arbitrary functions for which we only require the existence of the second derivative in space. 
In physical systems, the effect of the fields is to drag the densities of the two species in some direction (which depends on the function $\Psi_{1},\Psi_{2}$ themselves). As a consequence,  besides the diffusive flux, this drag creates an additional drift (sometimes called Ohmic) flux.  

We  assume that the system is coupled with the external environment via Dirichlet boundary conditions, fixing the densities for each species at $\xi=0$ and $\xi=\xi^{R}$. Namely, we endow the system \eqref{eqn:HD-equations LDP_grouped} with the following stationary boundary conditions valid for all $t\geq 0$
\begin{equation}\label{Eq:partialBC}
\begin{split}
	&\rho_{1}(0,t)=\rho_{1}^{L},\qquad \rho_{1}(\xi^{R},t)=\rho_{1}^{R},\\
	&\rho_{2}(0,t)=\rho_{2}^{L},\qquad \rho_{2}(\xi^{R},t)=\rho_{2}^{R}\,.
    \end{split}
\end{equation}
As a consequence of \eqref{exclusion-macro}, these boundary conditions satisfy 

\begin{equation}\label{Eq:ContraintsGlobalDensity}
	0\leq \rho_{1}^{L}+\rho_{2}^{L}\leq 1\quad \text{and}\quad 	0\leq \rho_{1}^{R}+\rho_{2}^{R}\leq 1\,.
\end{equation}

\subsection{Fluxes and continuity equations}
\label{SubSect:CurrentsAndConservationLaws}
To prepare for the study of uphill transport, we define the fluxes associated with the HDL model.
We denote $J_{i}$ the normalized flux of the $i$-th species crossing the infinitesimal volume surrounding a point $\xi\in [0,\xi^{R}]$ in an infinitesimal amount of time. 
It is usefull to rewrite the hydrodynamic equations as
\begin{equation}\label{eqn:HD-equations LDP}
\begin{split}
	\partial_{t}\rho_{1}
    & =\partial_{\xi\xi}\rho_{1}-\partial_{\xi} \left(2\rho_{1}\partial_{\xi} \Psi_1\right)+\partial_{\xi} \left(2\rho_{1}^2\partial_{\xi} \Psi_1 +
    2\rho_{1}\rho_{2}\partial_{\xi} \Psi_2\right)
    \ , \\
	\partial_{t}\rho_{2}
    & =\partial_{\xi\xi}\rho_{2}-\partial_{\xi} \left(2\rho_{2}\partial_{\xi} \Psi_2\right)+\partial_{\xi} \left(2\rho_{1}\rho_{2}\partial_{\xi} \Psi_1
    +2\rho_{2}^2\partial_{\xi} \Psi_2\right)
    \ .
    \end{split}
\end{equation}

Based on the RHS of Equations \eqref{eqn:HD-equations LDP}, we define four types of fluxes, depending on the physical phenomena that generate them. 

\begin{enumerate}
	\item \textit{Fick's (diffusion) fluxes} are generated by the concentration gradient. They read (first term on the RHS in \eqref{eqn:HD-equations LDP})
	\begin{equation}
		\begin{pmatrix}\label{eqn:fick-currents}
			J_{1}^{F}\\
			J_{2}^{F}
		\end{pmatrix}=-D(\rho_{1},\rho_{2})\begin{pmatrix}
			\partial_{\xi}\rho_{1}\\
			\partial_{\xi}\rho_{2}
		\end{pmatrix}=
		  \begin{pmatrix}
			1&0\\
			0&1
		\end{pmatrix}\begin{pmatrix}
			\partial_{\xi}\rho_{1}\\
			\partial_{\xi}\rho_{2} \ 
		\end{pmatrix},
	\end{equation}
where ${D}(\rho_{1},\rho_{2})$ is the so-called diffusion matrix, coincident with the identity matrix in this case. By consequence, Fick's fluxes are decoupled, meaning that the two species diffuse independently as a consequence of stirring.

	\item \textit{Drift fluxes} are generated by the interaction between one external field and the corresponding density. They read (second terms in Equations \eqref{eqn:HD-equations LDP})

	\begin{equation}\label{eqn:drift-currents}
		\begin{pmatrix}
			J_{1}^{D}\\
			J_{2}^{D}
		\end{pmatrix}=2\mathfrak{D}(\rho_{1},\rho_{2})\begin{pmatrix}
			\partial_{\xi} \Psi_{1}\\
			\partial_{\xi} \Psi_{2}
		\end{pmatrix}=2\begin{pmatrix}
			\rho_{1}&0\\
			0&\rho_{2}
		\end{pmatrix}\begin{pmatrix}
			\partial_{\xi} \Psi_{1}\\
			\partial_{\xi} \Psi_{2}
		\end{pmatrix}\, ,
	\end{equation}
where $\mathfrak{D}(\rho_{1},\rho_{2})$ is called \textit{drift} matrix. 

	\item \textit{Corrective fluxes} can be seen as alterations and interactions of the drift fluxes due to the competition among entities of the same or different species to occupy the available states under the action of the fields. They read (third terms in Equations \eqref{eqn:HD-equations LDP})
	\begin{equation}\label{eqn:exclusion-currents}
		\begin{pmatrix}
			J_{1}^{C}\\
			J_{2}^{C}
		\end{pmatrix}=2\mathcal{C}(\rho_{1},\rho_{2})\begin{pmatrix}
			\partial_{\xi} \Psi_{1}\\
			\partial_{\xi} \Psi_{2}
		\end{pmatrix}=2\begin{pmatrix}
			-\rho_{1}^2&-\rho_{1}\rho_{2}\\
			-\rho_{1}\rho_{2}&-\rho_{2}^2
		\end{pmatrix}\begin{pmatrix}
			\partial_{\xi} \Psi_{1}\\
			\partial_{\xi} \Psi_{2}
		\end{pmatrix}\,.
	\end{equation}

	Notice that the corrective fluxes are coupled (as the correction matrix has non-vanishing off-diagonal terms), non-linear, and, differently from the drift fluxes, quadratic in the densities. 
We also point out that the sum of the $\mathfrak{D}$ and $\mathcal{C}$ matrices, as defined above, coincides with the mobility matrix $\chi(\rho_{1},\rho_{2}$) defined in \cite{casini2025large} which reads
\begin{equation}\label{eqn:mobility-matrix}
		\chi(\rho_{1},\rho_{2})=\mathfrak{D}(\rho_{1},\rho_{2})+\mathcal{C}(\rho_{1},\rho_{2})=
        \begin{pmatrix}
			\rho_{1}(1-\rho_{1})&-\rho_{1}\rho_{2}\\
			-\rho_{1}\rho_{2}&\rho_{2}(1-\rho_{2})
		\end{pmatrix}
	\end{equation}
pointing out mutual exclusion as the physical origin of the corrective fluxes.
	\item \textit{Total fluxes}: they are given by the sum of Fick's, Drift, and Corrective fluxes for each species. Their expressions are:
	\begin{equation}\label{totalCurrent}
		\begin{pmatrix}
			J_{1}\\
			J_{2}
		\end{pmatrix}= \begin{pmatrix}
			J_{1}^{F}\\
			J_{2}^{F}
		\end{pmatrix}+ \begin{pmatrix}
			J_{1}^{D}\\
			J_{2}^{D}
		\end{pmatrix}+ \begin{pmatrix}
			J_{1}^{C}\\
			J_{2}^{C}
		\end{pmatrix}.
	\end{equation}
\end{enumerate}
Equations \eqref{totalCurrent} are routinely called transport equations. For each of the fluxes defined above, we further distinguish: 
\begin{itemize}
	\item \textit{Partial}: when we consider exclusively the i-th species. Namely, the partial total fluxes are  $J_{i}=J_{i}^{F}+J_{i}^{D}+J_{i}^{C}$ for $i\in \{1,2\}$.
	\item \textit{Global}: when we consider the sum over the species. Namely, $J=J_{1}+J_{2}$ is the global total flux, $J^F=J^F_{1}+J^F_{2}$ is the global diffusion flux, $J^F=J^F_{1}+J^F_{2}$ is the global drift flux, and $J^C=J^C_{1}+J^C_{2}$ is the global corrective flux.
\end{itemize}

If we enforce the functional dependence of $\Psi_1(\xi)$ and $\Psi_2(\xi)$ over the domain, and the boundary values $\rho_{1}^{L}$, $\rho_{2}^{L}$, $\rho_{1}^{R}$, $\rho_2^R$ then we can see the drift and corrective fluxes as \textit{bulk driven}, and the diffusion fluxes as \textit{boundary driven}. 
Notice, however, that in real-world systems of charged particles, the potentials $\Psi_1(\xi)$, $\Psi_2(\xi)$ are related, while $\rho_1$ and $\rho_2$ affect each other and the electric field via the Poisson equation, describing the system's electrostatics. 

This point is thoroughly addressed in Section \ref{Sect:Drift-diffusion models for charged particles}.  

Lastly, we notice that the HDL system \eqref{eqn:HD-equations LDP} can be obtained by substituting the total flux \eqref{totalCurrent} for each species in the respective continuity equation for the density 
	\begin{align}\label{Eq:Continuity}
		\partial_{t}\rho_{1}&=-\partial_{\xi} J_{1},\nonumber\\
		\partial_{t}\rho_{2}&=-\partial_{\xi} J_{2} \,.
	\end{align}
which expresses the conservation of the number of entities in the absence of reactions or other generation/annihilation processes.

\subsection{The steady state (SHDL)}
\label{SubSect:NESSUphill}
From now on, we focus on the so-called non-equilibrium steady state (sometimes abbreviated NESS) for the HDL equations \eqref{eqn:HD-equations LDP}. For a physical system, \textit{non-equilibrium} means that non-zero net fluxes exist with its surroundings, generating a natural tendency of the system to change its physical properties (composition, mass, charge,\ldots). 

{\em Steady state} simply means that all quantities ($\rho_{i},\Psi_{i}$ for $i\in \{1,2\}$) are stationary; that is, independent of time, so that all the time derivatives are zero, and the PDEs system \eqref{eqn:HD-equations LDP} reduces to a system of ODEs, hereafter denoted Stationary-HDL (SHDL)
\begin{subequations}\label{SHDL}
	\begin{align}
    &J_{1}=-\frac{d\rho_{1}}{d\xi}+2\rho_{1}(1-\rho_{1})\frac{d\Psi_{1}}{d\xi}-2\rho_{1}\rho_{2}\frac{d\Psi_{2}}{d\xi}\label{SHDL-c1}\\
		&J_{2}=-\frac{d\rho_{2}}{d\xi}+2\rho_{2}(1-\rho_{2})\frac{d\Psi_{2}}{d\xi}-2\rho_{1}\rho_{2}\frac{d\Psi_{1}}{d\xi}\label{SHDL-c2}\\
		&\frac{dJ_{1}}{d\xi}=0\label{SHDL-d1}\\
		&\frac{dJ_{2}}{d\xi}=0\ .\label{SHDL-d2}
	\end{align}
\end{subequations}

Once the SHDL system is solved, one can compute the Fick, drift, corrective, and total  fluxes in the non-equilibrium steady state by using expressions \eqref{eqn:fick-currents}, \eqref{eqn:drift-currents}, \eqref{eqn:exclusion-currents}, and \eqref{totalCurrent}, respectively.

The SHDL system of ODEs \eqref{SHDL} is driven out of equilibrium by the external fields generating drift fluxes, and/or by the diffusive force originating from the difference between the boundary density values. The net stationary fluxes stem from a competition between these two non-equilibrium sources. Since reactions are absent in this stationary formulation of the HDL model, Equations \eqref{Eq:Continuity} entail that fluxes are constant in space (Equations \eqref{SHDL-d1} and \eqref{SHDL-d2}). 

\subsection{Uphill transport for the SHDL model with constant drift}\label{Sect:UphillForSHDL}

In this section, firstly, we define the partial and the global uphill transport when the system is in the non-equilibrium
steady state and is endowed with Dirichlet boundary conditions. Then, we specialize to the case of linear potentials and move on to study uphill transport for the SHDL model \eqref{SHDL} by characterizing the uphill transport regimes.

\paragraph{Definition of uphill transport in the multispecies setup.}\label{Sect:DefinitionUphill} 
We consider the physical domain is the interval $[0,\xi^{R}]$ with $\xi^{R}>0$ and we introduce the global density $\rho:= \rho_{1}+\rho_{2}$, where the densities $\rho_{1},\rho_{2}$ solve the differential equations \eqref{SHDL}.
Moreover, it is convenient to introduce the global density at the boundaries
\begin{equation}
    \rho^{L}=\rho_{1}^{L}+\rho_{2}^{L}\leq 1 \quad \text{and}\quad  \rho^{R}=\rho_{1}^{R}+\rho_{2}^{R} \leq 1\,,
\end{equation}
and the differences of boundary densities
\begin{equation}
	\Delta_{1}=\rho_{1}^{R}-\rho_{1}^{L},\qquad \Delta_{2}=\rho_{2}^{R}-\rho_{2}^{L},\qquad \Delta=\Delta_{1}+\Delta_{2}=\rho^{R}-\rho^{L}\,.
\end{equation}
\begin{definition}[Uphill transport]\label{definition-uphill-diffusion}
In the non-equilibrium steady state, we define two types of uphill transport: 
\begin{itemize}
    \item \underline{Partial uphill transport.} We have partial uphill transport for the $i$-th species, with $i\in \{1,2\}$, when the $i$-th total flux $J_{i}$ has the same sign of $\Delta_{i}$, at each point of the spatial domain $[0,\xi^{R}]$.

    That is
    \begin{equation}
    \begin{split}
                \left(\min_{\xi\in [0,\xi^{R}]}J_{i}(\xi)\right)\times \Delta_{i} >0 \quad &\text{if}\quad \Delta_{i}>0\,,\\
                \left(\max_{\xi\in [0,\xi^{R}]}J_{i}(\xi)\right)\times \Delta_{i} >0 \quad &\text{if}\quad \Delta_{i}<0\,.
    \end{split}
    \end{equation}
    \item \underline{Global uphill transport.} We have global uphill transport when the global-total flux has the same sign of $\Delta$, on each point of the spatial domain $[0,\xi^{R}]$. 
    That is 
      \begin{equation}
      \begin{split}
       \left( \min_{\xi\in [0,\xi^{R}]}J(\xi)\right) \times \Delta>0\quad &\text{if}\quad \Delta>0\,,\\
       \left( \max_{\xi\in [0,\xi^{R}]}J(\xi)\right) \times \Delta>0\quad &\text{if}\quad \Delta<0.
       \end{split}
    \end{equation}
\end{itemize}
\end{definition}
Note that Definition \ref{definition-uphill-diffusion} is formulated having in mind a general case where reactions may be present and entities of the species can be generated or annihilated in the domain (as in \cite{casini2023uphill}). 
The definition is applicable also in our context; however, since reaction terms are absent, in the stationary state, the total fluxes are constant in space (because their divergence is zero) as per Equations  \eqref{SHDL-d1}-\eqref{SHDL-d2}. 

\paragraph{The case of a linear potential.}\label{Sect:CaseLinearPotential}

A convenient choice to investigate the possible occurence of uphill transport in the SHDL model is provided by linear potentials:
\begin{align}\label{potential-numerical}
    \Psi_{1}(\xi) = \frac{a}{2}\,\xi, 
    \qquad 
    \Psi_{2}(\xi) = \frac{b}{2}\,\xi,
\end{align}
where $a, b \in \mathbb{R}$ are independent of $\rho_{1}$ and $\rho_{2}$. Upon substitution into \eqref{eqn:HD-equations LDP}, the factors 2 cancel out and $a, b$ reveal their nature of constant field intensities associated with $\Psi_{1}$ and $\Psi_{2}$, respectively, which characterize uniform external driving forces imparted to the system. This linear choice significantly simplifies the analysis by reducing the effective parameter space of the model. Moreover, it provides a convenient benchmark for numerical studies and for direct comparison with the physical systems discussed in Section \ref{Sect:Drift-diffusion models for charged particles}.

Under the choice \eqref{potential-numerical}, the steady-state equations \eqref{SHDL} endowed with the boundary conditions \eqref{Eq:partialBC} reduce to a system of coupled second-order ordinary differential equations:
\begin{equation}\label{NESS-toSolve}
\begin{split}
    &\frac{d^{2}\rho_{1}}{d\xi^{2}}
    - a \left(\frac{d\rho_{1}}{d\xi} - 2\rho_{1}\frac{d\rho_{1}}{d\xi}\right)
    + b\left(\frac{d\rho_{1}}{d\xi}\rho_{2} + \rho_{1}\frac{d\rho_{2}}{d\xi}\right)
    = 0,  \\[0.5em]
    &\frac{d^{2}\rho_{2}}{d\xi^{2}}
    - b \left(\frac{d\rho_{2}}{d\xi} - 2\rho_{2}\frac{d\rho_{2}}{d\xi}\right)
    + a\left(\frac{d\rho_{1}}{d\xi}\rho_{2} + \rho_{1}\frac{d\rho_{2}}{d\xi}\right)
    = 0,  \\[0.5em]
    &\rho_{1}(0) = \rho_{1}^{L}, \quad 
     \rho_{1}(\xi^{R}) = \rho_{1}^{R}, \quad
     \rho_{2}(0) = \rho_{2}^{L}, \quad 
     \rho_{2}(\xi^{R}) = \rho_{2}^{R}.
     \end{split}
\end{equation}

The resulting model is fully determined by the two real parameters $a$ and $b$ and the four boundary values $\rho_{1}^L$, $\rho_{2}^L$, $\rho_{1}^R$, $\rho_{2}^R$. This parametrization allows for a systematic exploration of the interplay between drift and diffusive transport mechanisms in steady state.

In the following paragraph, we solve the boundary value problem \eqref{NESS-toSolve} by implementing a finite difference numerical method (FDM) with fourth-order accuracy, and obtain the simulation results reported below.
As an additional check, and for consistency with the developments in Section \ref{Sect: CaseStudy1}, we implemented Equations \eqref{NESS-toSolve} also in the general-purpose, finite elements (FEM) simulation framework FenicsX/Dolfinx \cite{baratta2023dolfinx}. In all tested cases, we found excellent agreement with the finite difference solution, thus providing strong, reassuring indications of a correct model implementation.

\paragraph{Uphill transport phase diagrams.}\label{sect:UphillPhaseDiagrams} 
In this paragraph, without loss of generality, we numerically solve the equations \eqref{NESS-toSolve} in the space interval $[0,1]$ (i.e., fixing $\xi^{R}=1$) to characterize the uphill transport regimes. 
Since the ODEs \eqref{NESS-toSolve} live in the 6-dimensional space of parameters spanned by the four boundary conditions \eqref{Eq:partialBC} subject to the constraint \eqref{Eq:ContraintsGlobalDensity} and the intensities of the external fields $a,b$, we proceed by fixing the boundary conditions for the two species, and then constructing phase diagrams by varying  $a,b\in [-2,2]$. For each value of the fields in this interval, we check whether the three total fluxes (one for each species and the global one) are directed uphill in the sense of Definition \ref{definition-uphill-diffusion}. 

Clearly, many setups can be constructed by choosing the boundary values. Here, we report two cases in which several types of uphill transport are observed. 

\paragraph{Simulation set \#1: $\Delta>0$, $\Delta_{1}>0$, $\Delta_{2}<0$ and $|\Delta_{1}|>|\Delta_{2}|$.}
Here, the gradients of species 1 and 2 imposed by the boundary conditions have opposite signs, and we look for the following types of uphill transport:
\begin{itemize}
	\item \textit{Partial uphill for the species 1:} since $\Delta_{1}>0$, we have uphill transport if $J_{1}>0$,
	\item \textit{Partial uphill for the species 2:} since $\Delta_{2}<0$, we have uphill transport if $J_{2}<0$,
	\item \textit{Global uphill:}
	since $\Delta>0$, we have uphill transport if $J>0$.
\end{itemize}

\begin{figure}[h]
\begin{subfigure}{0.5\textwidth}
\includegraphics[width=0.8\linewidth]{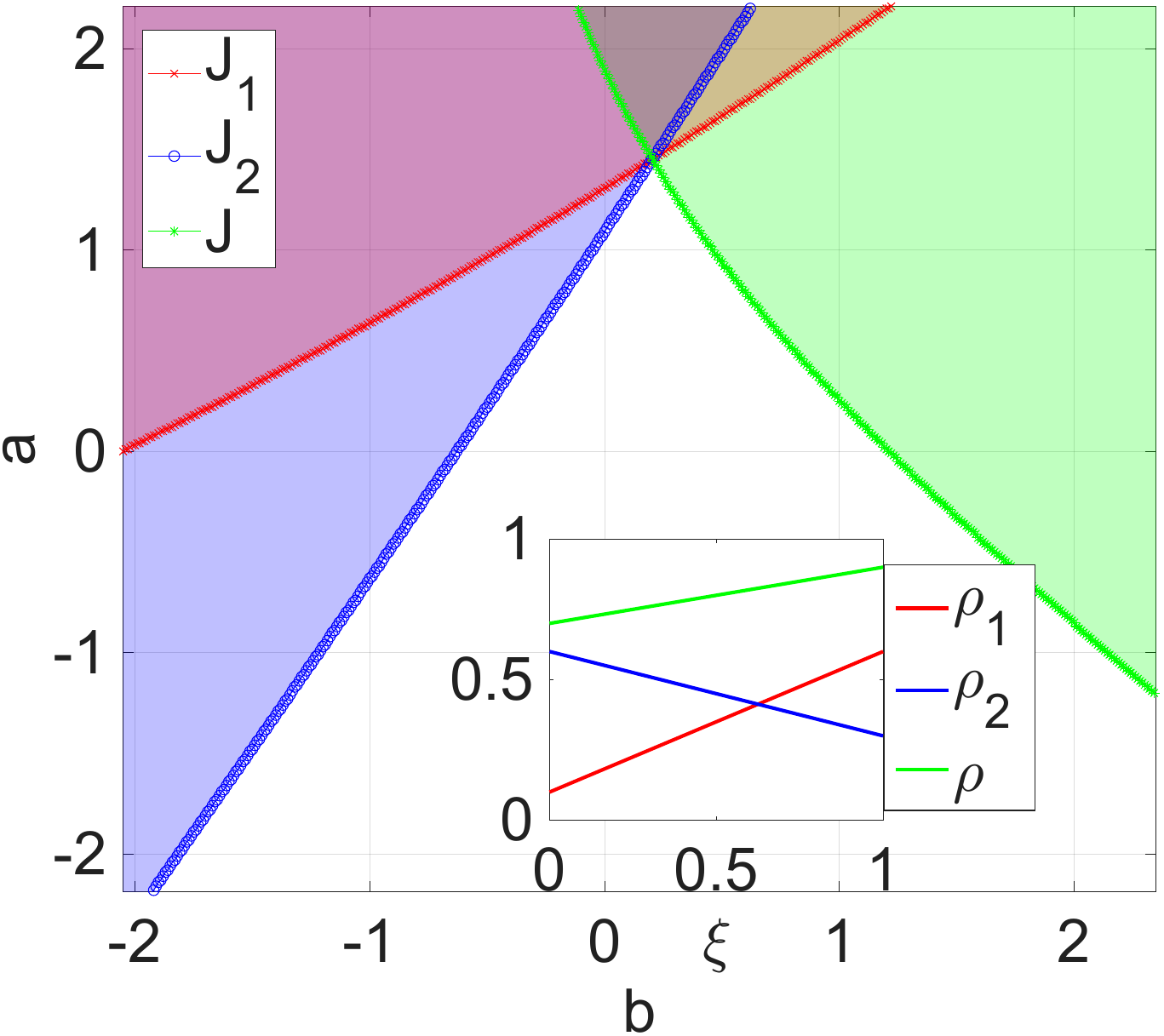}
	\centering\captionsetup{width=0.8\linewidth}
    \caption{Case without saturation at the boundaries. $\rho_{1}^{L}=0.1$, $\rho_{1}^{R}=0.6$, $\rho_{2}^{L}=0.6$, $\rho_{2}^{R}=0.3$, $\Delta_{1}=0.5$, $\Delta_{2}=-0.3$, $\Delta=0.2$.}
	\label{fig:fase 1-NOSAT}
\end{subfigure}
\begin{subfigure}{0.5\textwidth}
\centering\captionsetup{width=0.8\linewidth}
	\includegraphics[width=0.8\linewidth]{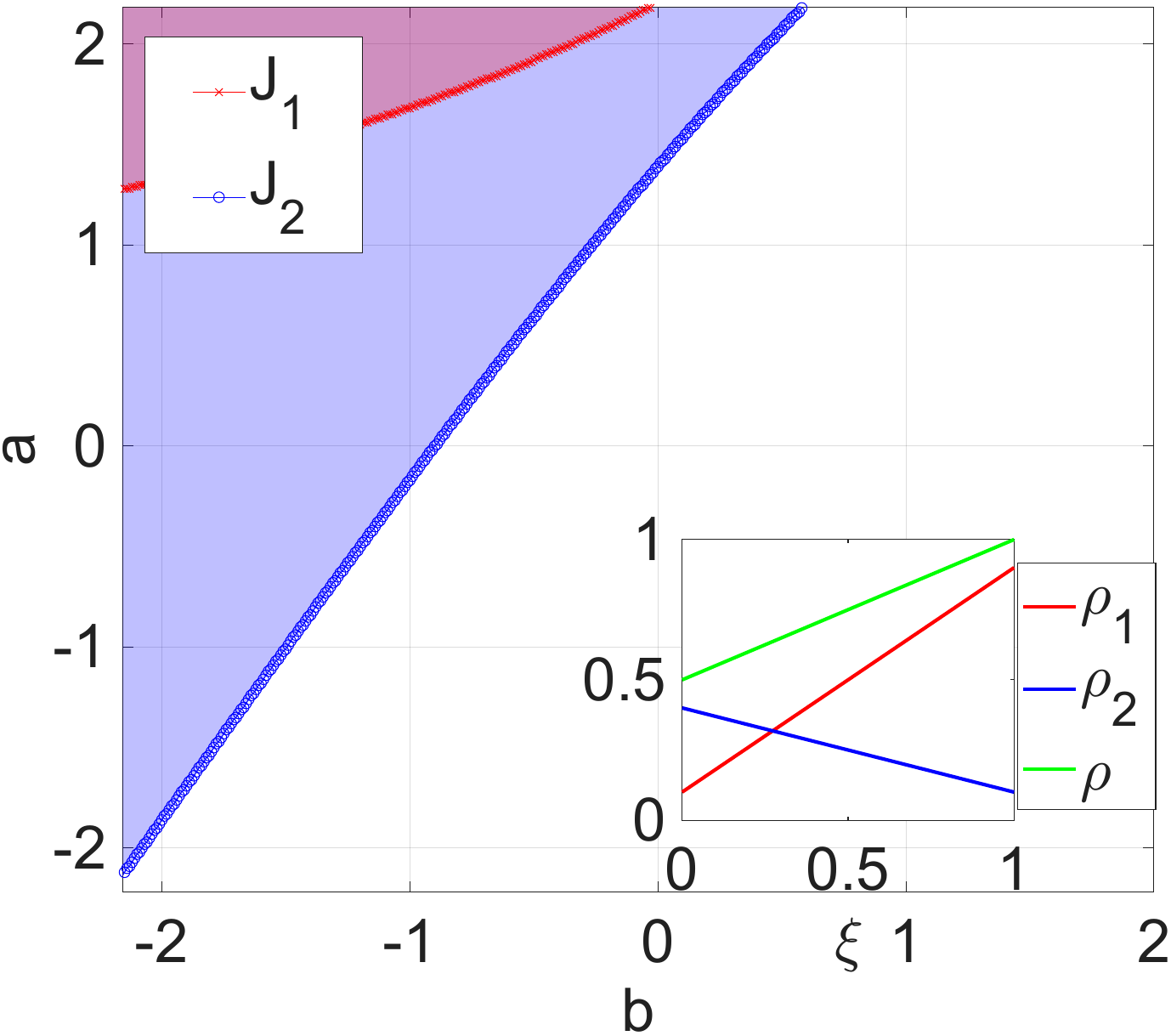}
	\caption{Case with saturation at the right boundary. $\rho_{1}^{L}=0.1$, $\rho_{1}^{R}=0.9$, $\rho_{2}^{L}=0.4$, $\rho_{2}^{R}=0.1$,   $\Delta_{1}=0.8$, $\Delta_{2}=-0.3$, $\Delta=0.5$.}
	\label{fig:fase 2-SAT}
\end{subfigure}
   \caption{
   The plots represent the lines of sign change (symbols), and the regions of uphill transport (shaded areas)  for the flux of species~1 (red), species~2 (blue), and for the global flux (green), in the space of the field parameters $a, b$. Notice the presence of regions where more than one uphill regime (partial or global) coexists. The insets show the linear interpolation of the boundary density values.}

\label{fig:image2}
\end{figure}

Figure \ref{fig:fase 1-NOSAT} (left plot) shows the simulation results for a set of boundary densities such that $\rho^{L}<1$ and $\rho^{R}<1$. 
In this case, we identify regions of the plot where: two or three types of uphill transport coexist; there is only one type of uphill; no uphill is visible.

Figure \ref{fig:fase 2-SAT} (right plot) shows the simulation results for the set of boundary densities where the gradient of $\rho_1$ increases (compared to Fig. \ref{fig:fase 1-NOSAT}), while that of $\rho_2$ stays the same, and $\rho^{R}=1$. Here, the phase diagram shows only three regions: a region where only partial uphill transport for species $2$ is possible; a region where the two partial uphills coexist; a region where no uphill is present. This result is easily interpreted considering that: 1) a larger Fick's flux  (compared to Fig. \ref{fig:fase 1-NOSAT}) is enforced for species 1 by the boundary conditions, making it more difficult to observe its partial uphill transport for the same range of fields; 2) since the global density at the boundary $\rho^{R}=1$, then $1-\rho_{1}-\rho_{2}$ vanishes. As a consequence, the global density no longer responds to the action of the external field, thereby impeding global uphill transport.
\paragraph{Simulation set \#2: $\Delta>0$, $\Delta_{1}>0$, $\Delta_{2}>0$ and $|\Delta_{1}|>|\Delta_{2}|$.}
Here, the gradients of species 1 and 2 imposed by the boundary conditions have the same signs, and we look for the following types of uphill transport:
\begin{itemize}
	\item \textit{Partial uphill for the species 1:} since $\Delta_{1}>0$, we have uphill transport if $J_{1}>0$,

	\item \textit{Partial uphill for the species 2:} since $\Delta_{2}>0$, we have uphill transport if $J_{2}>0$,

	\item \textit{Global uphill:}
	since $\Delta>0$, we have uphill transport if $J>0$.

\end{itemize}
\begin{figure}[ht]
\begin{subfigure}{0.5\textwidth}
	\includegraphics[width=0.8\linewidth]{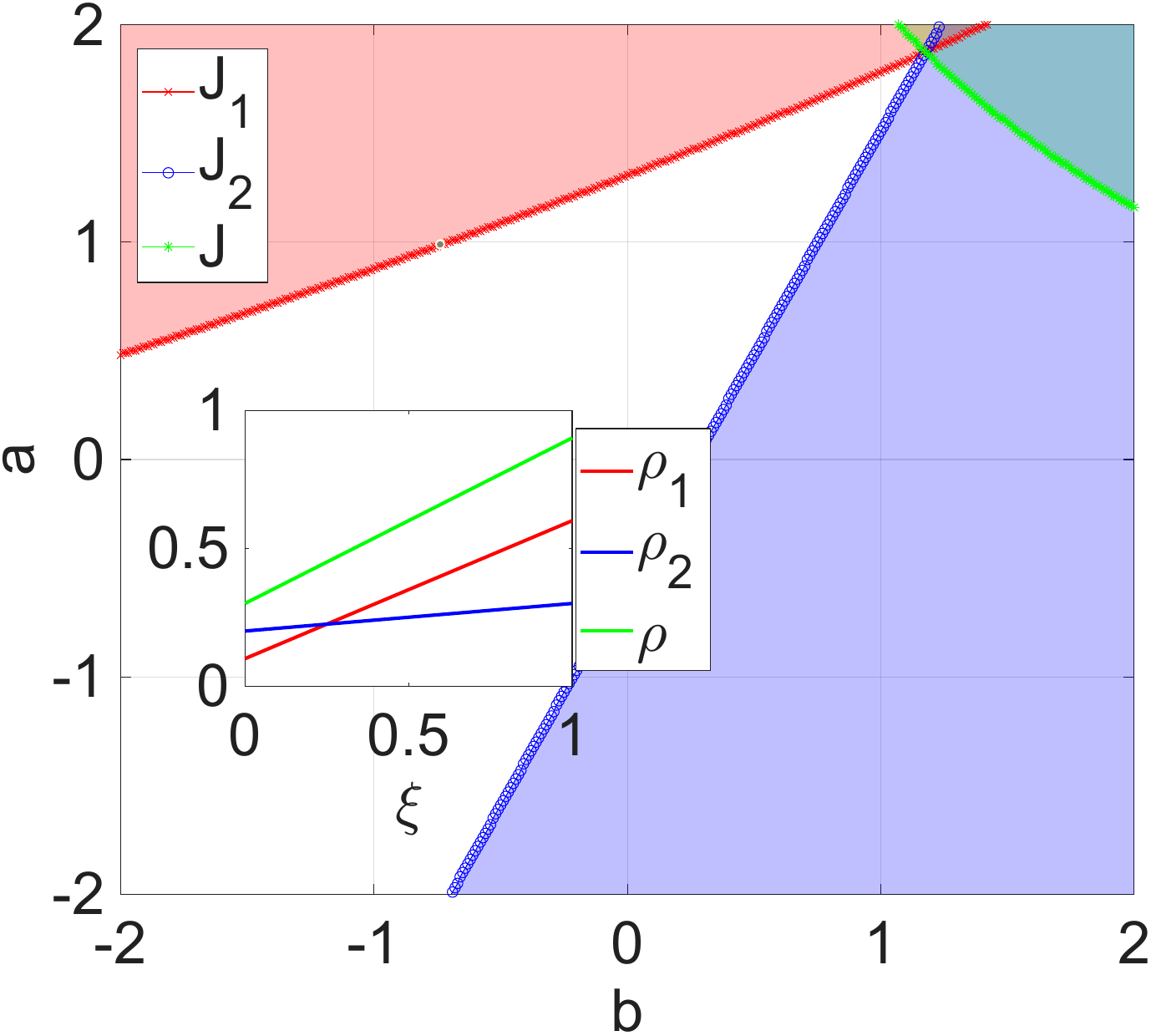}
	\centering\captionsetup{width=0.8\linewidth}
    \caption{Case without saturation at the boundaries. $\rho_{1}^{L}=0.1$, $\rho_{1}^{R}=0.6$, $\rho_{2}^{L}=0.2$, $\rho_{2}^{R}=0.3$,   $\Delta_{1}=0.5$, $\Delta_{2}=0.1$, $\Delta=0.6$}
	\label{fig:fase 3-NOSAT}
\end{subfigure}
\begin{subfigure}{0.5\textwidth}
\centering\captionsetup{width=0.8\linewidth}
	\includegraphics[width=0.8\linewidth]{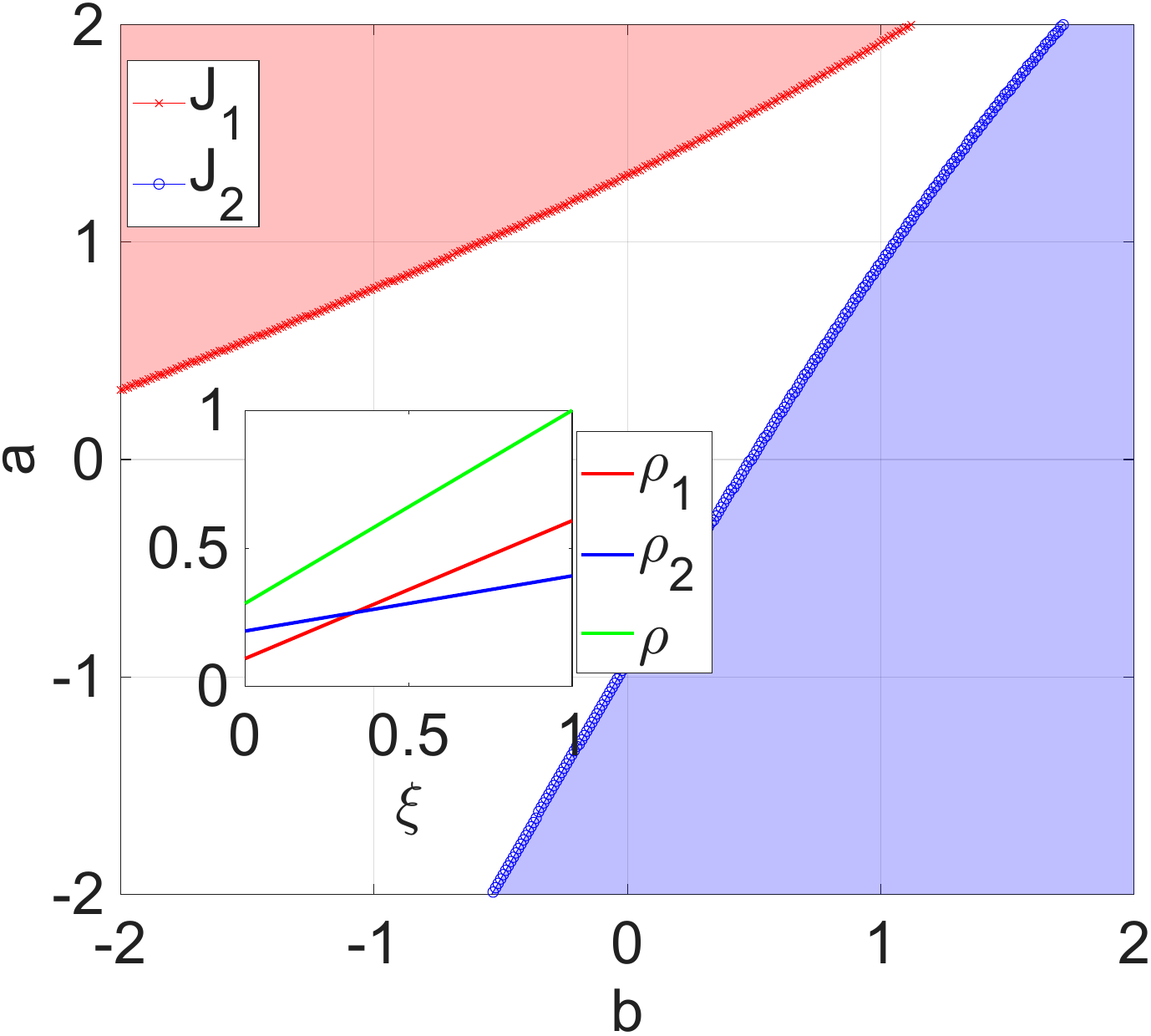}
	\caption{Case with saturation at the right boundary. $\rho_{1}^{L}=0.1$, $\rho_{1}^{R}=0.6$, $\rho_{2}^{L}=0.2$, $\rho_{2}^{R}=0.4$,   $\Delta_{1}=0.5$, $\Delta_{2}=0.2$, $\Delta=0.7$.}
\label{fig:fase 3-SAT}
\end{subfigure}
   \caption{
   The plots represent the lines of sign change (symbols), and the regions of uphill transport (shaded areas)  for the flux of species~1 (red), species~2 (blue), and for the global flux (green), in the space of the field parameters $a, b$. Notice the presence of regions where more than one uphill regime (partial or global) coexist in graph (a). The insets show the linear interpolation of the boundary density values.}
 	\label{fig:fase 3-SAT--}
\end{figure}

Figure \ref{fig:fase 3-NOSAT} shows the simulation results for a set of boundary densities such that $\rho^{L}<1$ and $\rho^{R}<1$. 
In this case, we identify regions of the plot where: two or three types of uphill transport coexist; there is only one type of uphill; no uphill is visible.  
Compared to Figure \ref{fig:fase 1-NOSAT}, the sign change in $\Delta_2$ swaps the position of the blue region below the line of sign change, as expected. 

Figure \ref{fig:fase 3-SAT} shows the simulation results for the set of boundary conditions in which $\rho^{R}=1$. Here, the phase diagram shows only three regions: one where only partial uphill for species $2$ is possible; one where only partial uphill for species $1$ is possible, and a last one without uphill. Similarly to Figure \ref{fig:fase 2-SAT}, no global uphill is possible in this case. Furthermore, as the gradient of $\rho_2$ is increased with respect to Figure \ref{fig:fase 3-NOSAT}, the region of partial uphill for species 2 shrinks.

Appendix \ref{appendix-A} reports a further simulation in which the boundary difference of global density is $\Delta=0$.  

\section{Drift-diffusion models for charged particles with volume exclusion}
\label{Sect:Drift-diffusion models for charged particles}

The HDL model defined by equations \eqref{eqn:HD-equations LDP}, and its uphill regimes, were derived by applying a mathematical procedure  (the hydrodynamic limit) to a microscopic model and then specializing to constant driving fields. In order to explore the applicability of the SHDL for studying uphill transport in systems of \textit{charged}, finite-size particles, we start focusing on a quite general template physical system whose essential dynamics shares the same governing transport mechanisms (drift and diffusion under volume exclusion constraints) as the microscopic model. 
Then, we introduce a set of commonly used model equations for that system (the so-called modified Poisson-Nernst-Plank model), and we unveil an important relation to the SHDL, which clarifies its underlying physical assumptions and validity limits.

\begin{figure}
    \centering
    \includegraphics[width=0.5\linewidth]{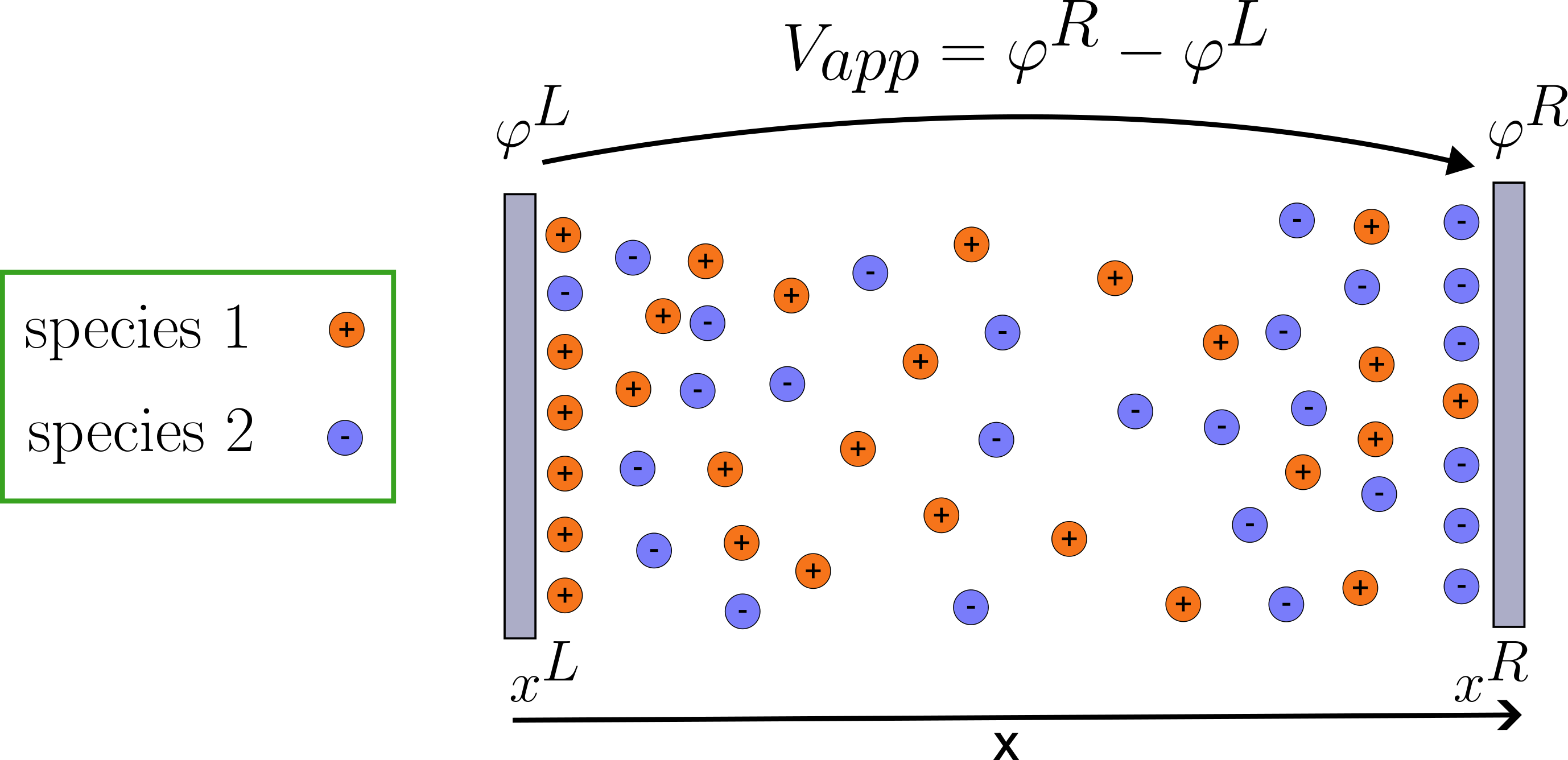}
    \caption{Sketch of the physical system that is mapped into the SHDL model. }
    \label{fig:basic_sketch}
\end{figure}

\subsection{The modified Poisson Nernst Planck model (mPNP)}
\label{subsect:Description}

Let's consider the one-dimensional domain in  Figure \ref{fig:basic_sketch} filled with a solvent-solute system immersed in the electric field generated by an electrostatic potential difference $V_{app}$, in Volt, between two ideal electrodes at the domain boundaries. The solute consists of charged particles with finite volume $d^3$ in m$^3$, bearing equal and opposite electrical charges (symmetric electrolyte), denoted by the unsigned valence $z = z_1 = -z_2$, with $z \in \mathbb{N}$ in units of the elementary charge q$\simeq$1.602$\cdot10^{-19}$ Coulomb. The species have volumetric, position dependent molar concentration $c_1(x), c_2(x)$ in units of mol/m$^3$; the position-dependent electrostatic potential $\varphi(x)$, in Volts, describes the electric field as ${E}=-\partial_\xi\,\varphi$ in V/m. 

Drift diffusion transport in systems as the one in Figure \ref{fig:basic_sketch} can be modeled with the so-called modified Poisson-Nernst-Planck model (mPNP), which is often used in scientific and engineering studies to predict the electrical responses of electrode/electrolyte junctions in exclusion conditions; for instance, in neuron/electrolyte clefts \cite{leva2024}, porous electrodes \cite{yang2022direct}, and fuel cell membranes \cite{de2025generalized}. 

The mPNP model has been derived in \cite{kilic2007,borukhov2000adsorption}, using the mean field approximation and free energy arguments, from a microscopic dynamics governed by an exclusion constraint where at most one particle may occupy a given site. 
Unlike the microscopic multispecies, weakly asymmetric stirring process introduced in Section \ref{subsect:microscopic-model}, the mPNP does not track the discrete dynamics of particles. Nonetheless, it embodies finite-size (so-called \emph{steric}) effects and accounts for the average effects of the finite particle diameter in the continuum.

The mPNP model for the steady state consists of the following set of equations \cite{kilic2007}:
\begin{subequations}\label{mPNP-NoNorm}
\begin{align}
    \Phi_1 &= -D_1\frac{dc_1}{dx} - z\mu_1c_1\frac{d\varphi}{dx} - D_1 c_1 \frac{N_Ad^3\big(\frac{dc_1}{dx} + \frac{dc_2}{dx}\big)}{1-N_Ad^3\big(c_1+c_2\big)}\label{mPNP-NoNorm-c1}\\
    \Phi_2 &= -D_2\frac{dc_2}{dx} + z\mu_2c_2\frac{d\varphi}{dx}- D_2 c_2 \frac{N_Ad^3\big(\frac{dc_1}{dx} + \frac{dc_2}{dx}\big)}{1-N_Ad^3\big(c_1+c_2\big)}\label{mPNP-NoNorm-c2}\\
    \frac{d\Phi_1}{dx}&=0\label{mPNP-NoNorm-d1}\\
    \frac{d\Phi_2}{dx}&=0\label{mPNP-NoNorm-d2}\\
        -\varepsilon\frac{d^2\varphi}{dx^2} &= zF(c_1 -c_2) \label{mPNP-NoNorm-Poiss}
\end{align}
\end{subequations}
where $\varepsilon$ is the dielectric permittivity of the electrolyte, $F=qN_A$ is Faraday's constant, $N_A$ is Avogadro's number. 
For $i\in \{1,2\}$, $D_i$ is the diffusivity constant, $\Phi_i$ is the flux per mole of particles and $\mu_i$ is the mobility. 

Equations \eqref{mPNP-NoNorm-c1} and \eqref{mPNP-NoNorm-c2} are the drift diffusion transport equations, which play in the mPNP a similar role as Equations \eqref{SHDL-c1} and \eqref{SHDL-c2} in the SHDL model. The first two terms at the right-hand side play the same role as $J^F_i$ and $J^D_i$. The third term has the same role as the $J^C_i$ in Equation \eqref{eqn:exclusion-currents}; i.e., it accounts for the steric exclusion due to the finite particle size. In fact, it cross-couples the fluxes and densities of the species, and vanishes for $d\rightarrow 0$. 
The limit of the mPNP for $d\rightarrow 0$ is known as Poisson-Nernst-Planck (PNP) or Poisson-Drift-Diffusion model, and can predict uphill transport only in the trivial case where $|J_i^D|>|J_i^F|$.
Steric (exclusion) effects are significant when the local molar concentrations $c_{i}$ approach the close-packing limit of order $1/(N_A d^{3})$. Notice that, differently from Equations \eqref{SHDL-c1}, \eqref{SHDL-c2}, these terms depend on the concentration gradients, not on the potential gradients.

Compared to the SHDL, the Poisson Equation \eqref{mPNP-NoNorm-Poiss} for a uniform isotropic medium is added to the transport and continuity equations to self-consistently account for electrostatic interactions among the charged particles' distributions and the electric field. In this physical context, Equation \eqref{mPNP-NoNorm-Poiss} generates an additional coupling between the densities and the potential besides the one represented by the $J^C_i$ corrective terms. Notice that distributed reaction or generation/annihilation terms, not considered here, could also couple the transport equations in ways that could affect uphill transport phenomena \cite{casini2023uphill}.

\subsection{Normalization of the mPNP model}
\label{subSect:normalization}

To obtain unitless variables matching the ones of the hydrodynamic model in Section \ref{Sect:UphillPDEs}, and to ease the later solution of the equations in the Fenicsx/Dolfinx environment \cite{baratta2023dolfinx}, we apply the following normalizations:

\begin{equation}\label{eq:Normalizations}
    \begin{cases}
    c_1 & = c_{norm}\, \rho_1 = \frac{1}{N_Ad^3} \ \rho_1 \\
    c_2 & = c_{norm}\, \rho_2 = \frac{1}{N_Ad^3} \ \rho_2 \\
    \varphi & = \varphi_{norm}\, \Psi= \frac{RT}{F} \ \Psi = \frac{k_BT}{q} \ \Psi\\
    x & = x_{norm} \, \xi = \sqrt{\frac{\varepsilon kT}{qFc_{norm}}} \ \xi \\
    \Phi_i &= \Phi_{norm}\, J_i=(D_i \, c_{norm}/{x_{norm}})\, J_i\,, 
\end{cases}
\end{equation}
where Einstein's relation entails $\mu_i=({F/RT}){D_i}=(q/k_BT){D_i}$, $T$ is the absolute temperature, $R$ is the ideal gas constant and $k_B$ is Boltzmann constant.
Here, $\rho_1$ and $\rho_2$ represent the normalized, unitless concentrations; they are both $\leq 1$ due to the exclusion requirement. $\Psi$ is the normalized, unitless potential, $c_{norm}, \varphi_{norm}, x_{norm}$, and $\Phi_{norm}$ are the normalization constants.

Thus, the transformed ODEs for the steady state of the mPNP model \eqref{mPNP-NoNorm} become:

\begin{subequations}\label{mPNP}
\begin{align}
    &J_1 = -\frac{d\rho_1}{d\xi} - z\rho_1\frac{d\Psi}{d\xi} -\frac{\rho_1}{1-\rho_1-\rho_2}\left (\frac{d\rho_1}{d\xi} + \frac{d\rho_2}{d\xi}\right )\label{mPNP-c1} \\
    &J_2 = -\frac{d\rho_2}{d\xi} + z \rho_2\frac{d\Psi}{d\xi} -\frac{\rho_2}{1-\rho_1-\rho_2}\left (\frac{d\rho_1}{d\xi} + \frac{d\rho_2}{d\xi}\right )\label{mPNP-c2} \\
    &\frac{dJ_1}{d\xi} = 0 \label{mPNP-d1}\\
    &\frac{dJ_2}{d\xi} = 0\label{mPNP-d2}\\
    &\frac{d^2\Psi}{d\xi^2} = - z(\rho_1-\rho_2)\label{mPNP-Poisson}\,.
    \end{align}
\end{subequations}

\subsection{The self-consistent SHDL model (P-SHDL)}
\label{SubSect:matching}

Taking inspiration from the mPNP model discussed above, we aim at developing a version of the SHDL model where particles possess an electrical charge. Charges respond to electric fields, and the field, described by the normalized electrostatic potential $\Psi(x)$, must act with opposite sign on particles of opposite charge. In the SHDL model, we can achieve this by setting
\begin{equation}\label{potential-assumptions}
    \Psi_{2}(\xi)=-\Psi_{1}(\xi)=\Psi(\xi)\qquad \forall \xi \in[0,1] \, .
\end{equation}
Moreover, the potential cannot be chosen arbitrarily but has to satisfy the Poisson equation \eqref{mPNP-Poisson}.
We thus obtain a self-consistent version of the SHDL model, hereafter denoted Poisson-SHDL (P-SHDL) defined by the equations
\begin{subequations}\label{P-SHDL}
	\begin{align}
	    &J_1 = -\frac{d\rho_1}{d\xi} - z\rho_1\frac{d\Psi}{d\xi} -z\rho_1(\rho_2-\rho_1)\frac{d\Psi}{d\xi}\label{P-SHDL-c1}\\ 
    	&J_2 = -\frac{d\rho_2}{d\xi} + z\rho_2\frac{d\Psi}{d\xi} -z\rho_2(\rho_2-\rho_1)\frac{d\Psi}{d\xi}\label{P-SHDL-c2}\\
		&\frac{dJ_1}{d\xi} = 0\label{P-SHDL-d1} \\
		&\frac{dJ_2}{d\xi} = 0 \label{P-SHDL-d2}\\
         &\frac{d^2\Psi}{d\xi^2} = - z(\rho_1-\rho_2)\label{P-SHDL-Poisson}\,.
	\end{align}
\end{subequations}
We observe that for $z=2$, the above equations may be viewed as the SHDL model \eqref{SHDL} where the potentials are chosen as 
in \eqref{potential-assumptions}
and where the Poisson equation has been added.

Given the density profiles $\rho_{1}(\xi)$ and $\rho_{2}(\xi)$, and the Dirichlet boundary conditions 
\begin{equation}
    \begin{split}
        \Psi^{L} & = 0\quad \text{and}\quad 
            \Psi^{R} \in \mathbb{R}\, , \label{BC-P-SHDL-Psi}
    \end{split}
\end{equation}
the solution of the second-order Poisson equation \eqref{mPNP-Poisson} uniquely determines the potential. Indeed, based on the Leibnitz integral rule, straightforward calculations lead to the expression:  

\begin{equation}\label{soluzione}
    \Psi(\xi)
    = \frac{\xi}{\xi^{R}} \left [ \Psi^{R}
    - z \int_{0}^{\xi^{R}} (\xi^{R} - y)\, [\rho_{2}(y) - \rho_{1}(y)]\, dy \right ]
    + z\int_{0}^{\xi} (\xi - y)\, [\rho_{2}(y) - \rho_{1}(y)]\, dy
    \ \ .
\end{equation}

Equation \eqref{soluzione} puts in evidence that $\Psi(\xi)$ is nonlinear in $\xi$ due to the last integral on the RHS. 
However, for sufficiently short domains on the scale of the normalization length \eqref{eq:Normalizations}, or for small density values (the so-called dilute species regime), or for small total charge density $(\rho_{2}(y) - \rho_{1}(y))$, the integral is very small, so that deviations from linearity remain negligible, the potential can be well approximated by the first (linear) term on the RHS only, and the field is constant (as assumed in the SHDL).

\subsection{The P-SHDL model as near-equilibrium approximation of the mPNP.} 
\label{subsect:The P-SHDL model as near-equilibrium}
If we impose equilibrium, the  mPNP system~\eqref{mPNP} reduces to the so called modified Poisson-Boltzmann model (mPB) derived in \cite{borukhov2000adsorption} both from a discrete-lattice approach, and from a continuum approach, minimizing the total energy of the system. In the literature the use of the mPB model is widespread \cite{anwar2024salinity}\cite{abdel2023volume}\cite{garza2018mechanism} to describe a variety of physical systems as the one in Figure \ref{fig:basic_sketch}. The mPB yields explicit expressions for the concentrations as a function of potential in exclusion conditions (Equation (18) in \cite{borukhov2000adsorption}), here reported for convenience

\begin{equation}
    c_1 = \frac{c_1^{*}\exp\big(-\frac{zF(\varphi-\varphi^{*})}{RT}\big)}{1 - 2N_Ac_1^{*}d^3 + 2N_Ad^3c_1^{*} \cosh \big(\frac{zF(\varphi-\varphi^{*})}{RT}\big)}\label{eqn:equilibrium_mpnp1}
\end{equation}
\begin{equation}
    c_2 = \frac{c_2^{*}\exp\big(\frac{zF(\varphi-\varphi^{*})}{RT}\big)}{1 - 2N_Ac_2^{*}d^3 + 2N_Ad^3c_2^{*} \cosh \big(\frac{zF(\varphi-\varphi^{*})}{RT}\big)},\label{eqn:equilibrium_mpnp2}
\end{equation}
where $c_1^{*}$, $c_2^{*}$, and $\varphi^{*}$ are the species' concentrations and electrostatic potential at a reference location. If we impose charge neutrality and $\varphi^*=0\,$V at the reference location, then the reference concentrations coincide, and we find 
\begin{equation}
    c_1 = \frac{c^{*}\exp\big(-\frac{zF\varphi}{RT}\big)}{1 - 2N_Ac^{*}d^3 + 2N_Ad^3c^{*} \cosh \big(\frac{zF\varphi}{RT}\big)}\label{eqn:equilibrium_mpnp1_charge_neutrality}
\end{equation}
\begin{equation}
    c_2 = \frac{c^{*}\exp\big(\frac{zF\varphi}{RT}\big)}{1 - 2N_Ac^{*}d^3 + 2N_Ad^3c^{*} \cosh \big(\frac{zF\varphi}{RT}\big)}.\label{eqn:equilibrium_mpnp2_charge_neutrality}
\end{equation}
Upon normalization according to Equations \eqref{eq:Normalizations}, and provided 
\begin{equation}
    \rho_{1}^{L} = \rho_{2}^{L} = \rho^{*} \leq 1/2\, ,  
    \label{BC-P-SHDL-Rho}
\end{equation}
so that $\rho^L\leq1$, we obtain
\begin{equation}
    	\rho_1(\xi) = \frac{\exp (-z\Psi(\xi))}{2\cosh (z\Psi(\xi)) + (\frac{1}{\rho^{*}}-2)}, \qquad \rho_2(\xi) = \frac{\exp (z\Psi(\xi))}{2\cosh (z\Psi(\xi)) + (\frac{1}{\rho^{*}}-2)} \ \ \ \ . \label{eqn:equilibrium_mpnp_scaled1}
\end{equation}

A direct calculation shows that if we evaluate the sum of the concentration gradients with the \textit{equilibrium} expressions \eqref{eqn:equilibrium_mpnp_scaled1} we get
\begin{align}\label{eq:SumOfRhoDerivatives}
	\frac{1}{1-\rho_2-\rho_1}\left (\frac{d\rho_1}{d\xi} + \frac{d\rho_2}{d\xi}\right ) &= \frac{1}{1-\rho_2-\rho_1}z \cdot\frac{(\frac{1}{\rho^{*}}-2) \left[\exp (z\Psi) - \exp (-z\Psi)\right]}{\big(2\cosh (z\Psi) + (\frac{1}{\rho^{*}}-2)\big)^2}\cdot \frac{d\Psi}{d\xi} \nonumber\\
    &= \frac{z(\rho_2-\rho_1)(1-\rho_2-\rho_1)}{1-\rho_2-\rho_1}\cdot \frac{d\Psi}{d\xi} =  z(\rho_2-\rho_1)\cdot \frac{d\Psi}{d\xi}\ \ \ \ .
\end{align}
By comparing the first and the last term of the above equations to the $J^C_i$ terms of the expressions \eqref{mPNP} and \eqref{P-SHDL}, respectively, 
we realize that the P-SHDL is an mPNP model in which the steric correction term has been estimated with the equilibrium expressions of the charge density profiles, Equations \eqref{eqn:equilibrium_mpnp_scaled1}. This result clarifies the physical grounds of the SHDL model and relates it to the mPNP and mPB models in use in the engineering community. 

\section{Models' comparison}\label{Section-comparison-OutOfEq}
In this section, we compare via numerical simulations the three models introduced so far: SHDL, mPNP, and P-SHDL. The comparison is carried out in general, out-of-equilibrium setups, achieved by imposing Dirichlet boundary conditions on the right edge of the domain to generate non-vanishing fluxes. Namely, we freely chose $\rho_1^R$ and $\rho_2^R$ with the only constraint that 
\begin{equation}
    0 \leq \rho_{1}^{R} + \rho_{2}^{R} = \rho^R \leq 1 \ \ .
\end{equation}
To ensure a consistent comparison in view of the uphill transport studies in Section\,\ref{Sect:Uphill-Section}, we also impose the boundary conditions \eqref{BC-P-SHDL-Psi} and \eqref{BC-P-SHDL-Rho} on the model equations, and the field is constant in SHDL calculations. 
Numerical experiments were conducted assuming $z=2$, $d=0.72$ nm, consistently with \cite{nightingale1959phenomenological}, i.e., $1/N_Ad^3\simeq$4.5 mol/l, $D_1=D_2 =1\cdot 10^{-9} $m$^2$/s for a domain length $x^R= 10\,$nm equal to $\xi^{R} \simeq 48$ normalization lengths according to Equation \eqref{eq:Normalizations}). 
We distinguish \emph{low-} and \emph{high-voltage} regimes, according to $\varphi^R$ being in the few tens and few hundreds of millivolts, respectively. These voltages correspond to electric fields ranging from tens to hundreds of kV/cm, respectively. These field values approach the limits at which the viscoelectric effect \cite{hunter2013zeta} and the dielectric saturation of water \cite{booth1951dielectric} become important.
In addition, we explored both \emph{low-} and \emph{high-concentration} regimes, corresponding to normalized boundary density values of order $10^{-2}$ and $10^{-1}$, respectively, as detailed in the figures and captions.
\begin{figure}[ht!]
    \centering
    \includegraphics[width=0.75\linewidth]{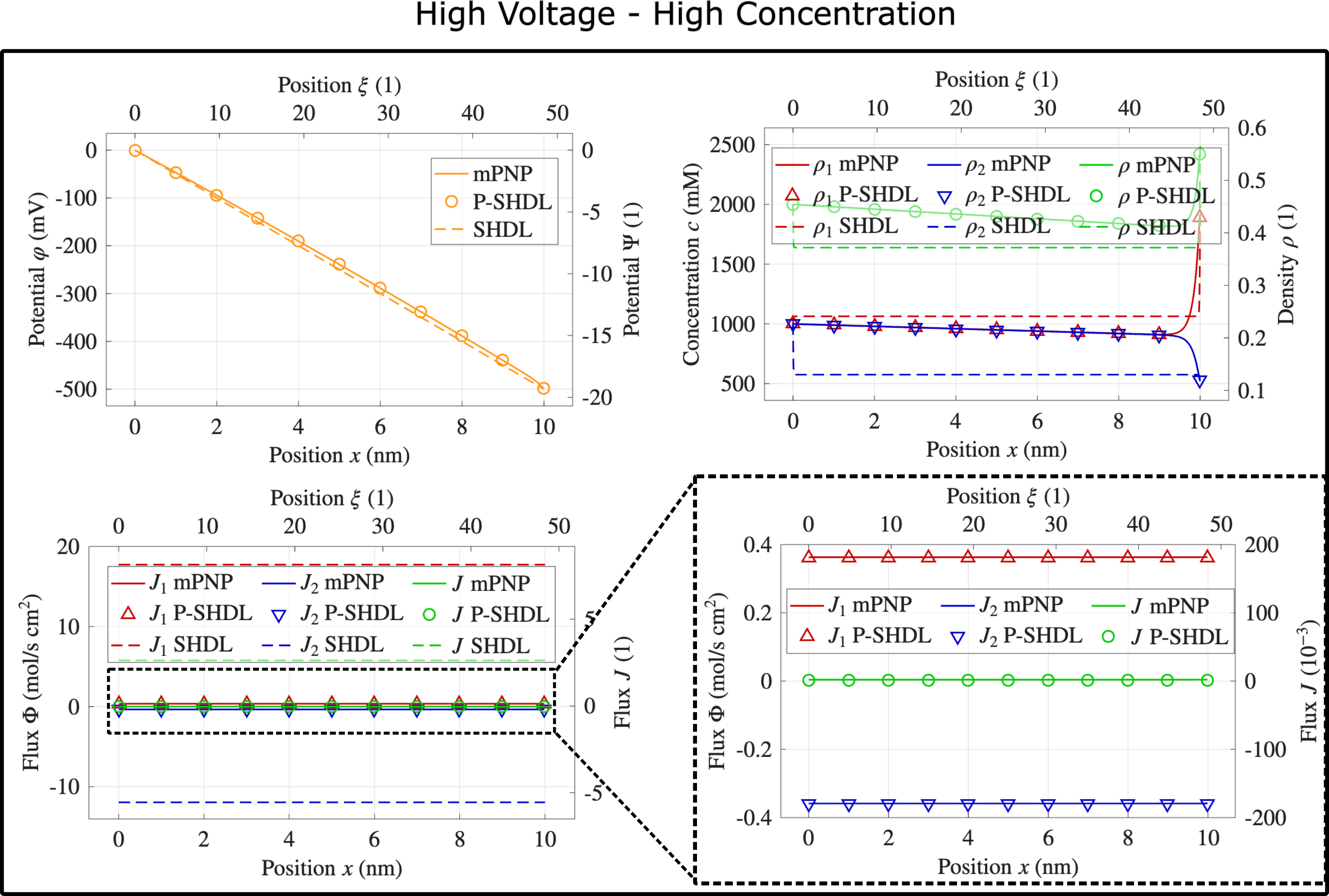}
    \caption{Comparison of the mPNP, P-SHDL, and SHDL models at {high voltage and high boundary densities}. 
    The mPNP and P-SHDL profiles nearly coincide, confirming the validity of the P-SHDL equilibrium approximation in this parameters' range. 
    The SHDL model deviates from the other two models, reflecting the violation from the Poisson equation and the linear potential profile. Nonetheless, the fluxes of species $1$ and $2$ have the same sign for the three models. Boundary conditions: $\Psi^R = 19.34$, $\Psi^L = 0$, $\rho_1^R=0.45$, $\rho_2^R=0.11$, $\rho_1^L=\rho_2^L=0.22$, corresponding to $\varphi^R = 500\,$mV, $\varphi^L = 0\,$mV, $c_1^R=2000\,$mM, $c_2^R=500\,$mM, $c_1^L=c_2^L=1000\,$mM respectively.}
    \label{fig:HV-HC}
\end{figure}

\begin{figure}[ht!]
    \centering
    \includegraphics[width=0.75\linewidth]{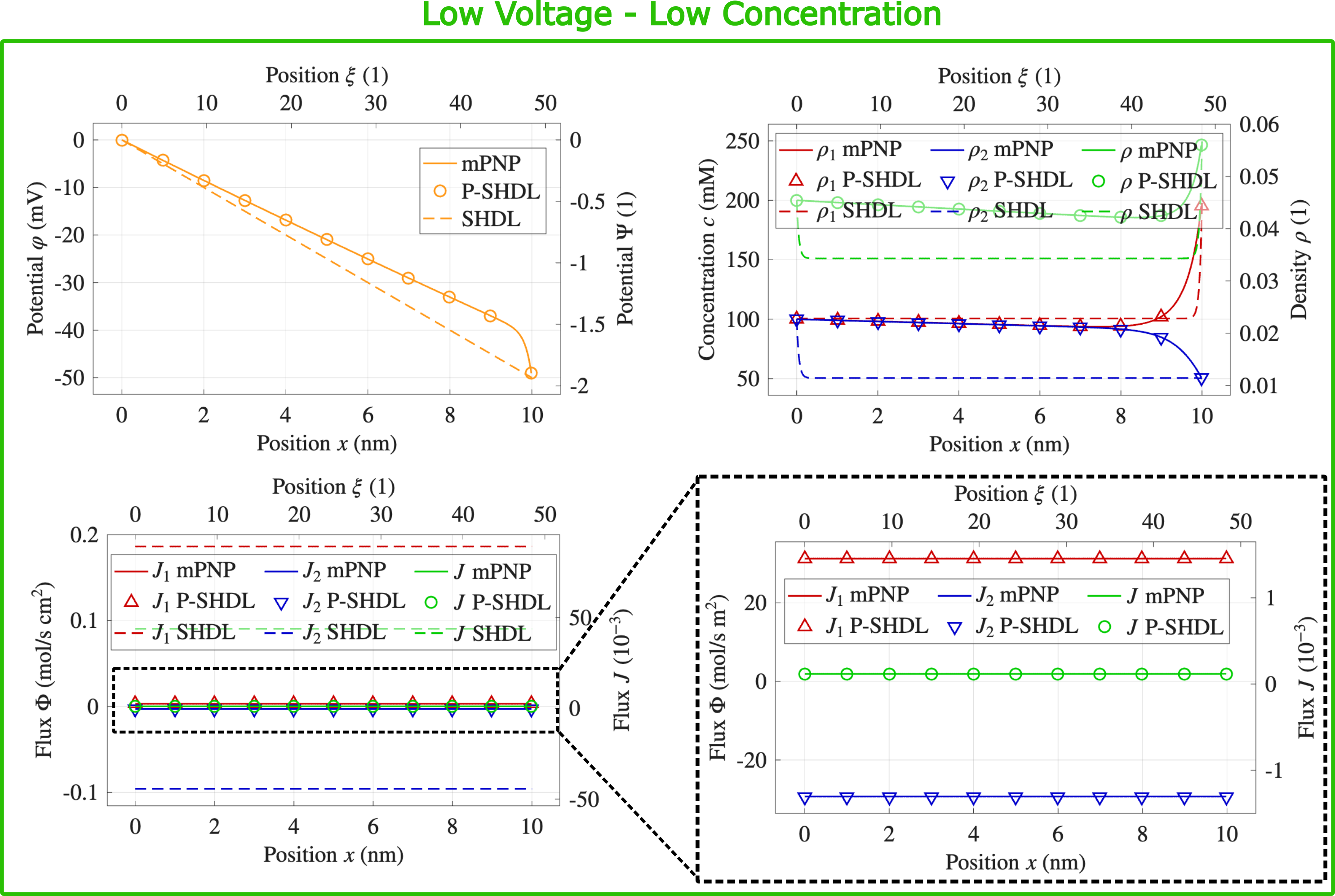}
    \caption{Comparison of the mPNP, P-SHDL, and SHDL models at {low voltage and low boundary densities}.  
    The mPNP and P-SHDL profiles overlap closely, while the SHDL model shows divergence in $\rho_{2}(\xi)$, but a good match with mPNP and P-SHDL for the profile $\rho_{1}(\xi)$. 
    Nevertheless, the potential deviates from linearity, illustrating the residual effect of the Poisson equation. Nonetheless, the fluxes of species $1$ and $2$ have the same sign for the three models. Boundary conditions: $\Psi^R = 1.934$, $\Psi^L = 0$, $\rho_1^R=0.045$, $\rho_2^R=0.011$, $\rho_1^L=\rho_2^L=0.022$, corresponding to $\varphi^R = 50\,$mV, $\varphi^L = 0\,$mV, $c_1^R=200\,$mM, $c_2^R=50\,$mM, $c_1^L=c_2^L=100\,$mM respectively.}
    \label{fig:LV-LC}
\end{figure}

Figures~\ref{fig:HV-HC} (high-voltage, high-concentration) and \ref{fig:LV-LC} (low-voltage, low-concentration) present a systematic comparison among the mPNP, P-SHDL, and SHDL models across the voltage and boundary density regimes above. All quantities are given in both physical (left y-axis/bottom x-axis) and normalized units (right y-axis/top $\xi$-axis). 

We highlight the following observations. 
\begin{enumerate}
    \item In both cases, the mPNP and P-SHDL \textit{potential} (top left graphs), and \textit{density} (top right graphs) are nearly coincident. This suggests that in the explored parameter range, near-equilibrium conditions hold. 
    \item In both cases and for the mPNP and P-SHDL  models, the difference between the partial \textit{density} profiles (top right graphs) starts from zero at the left boundary, as per the boundary condition, and stays very small across the domain, except near the right contact where a net charge ($\rho_1(\xi)-\rho_2(\xi)$) buildup is needed to satisfy the boundary condition consistently with the Poisson equation. 
    For the SHDL model, where the Poisson equation is not solved, the particle density is remarkably different from the other models (except at the left boundary), although still within about a factor of two from it. This suggests that the electrostatic interaction between potential and charges expressed by the Poisson equation tends to maintain a zero net charge across most of the domain, whose absence in the SHDL model impacts the potential profile as discussed below.
    \item For the SHDL model the \textit{potential} is linear by hypothesis. 
    For the mPNP and P-SHDL models, as a consequence of the small net charge density, $\rho_1(\xi)-\rho_2(\xi)$, and the Poisson equation, the \textit{potential} profiles (top left graphs) are nearly linear, with modest deviations 
    from the SHDL profile, only visible at low voltage due to the expanded graph scales. This justifies the limited discrepancy among the models observed above in the charge density profiles.
    \item In both figures and for all models, the domain is not globally charge neutral; that is, the integral of the RHS of the Poisson equation over the domain is not zero. This is due to the non-zero net charge and fluxes imposed at the right boundary.
    \item In both cases and for all models, the \textit{flux} profiles are constant in space (bottom left graphs), as expected in the absence of reaction or other generation/annihilation terms, equations \eqref{mPNP-d1}, \eqref{mPNP-d2}, \eqref{SHDL-d1}, and \eqref{SHDL-d2}. 
    \item In both cases, the \textit{flux} profiles for the mPNP and P-SHDL models (bottom graphs) are nearly coincident, consistent with the agreement among the potential and density profiles, point 1 of this list. 
    \item The amplitude of the partial and global \textit{fluxes} is quite small in the mPNP and P-SHDL models (bottom right graphs) where electrostatic coupling tends to minimize the perturbation due to the flux. It is much larger in the SHDL model (bottom left graphs). However, the sign of all the SHDL fluxes is the same as for the mPNP and P-SHDL models. Following the definitions in Section \ref{Sect:DefinitionUphill}, this means that all models lead to the same predictions of uphill transport regimes in the explored parameter ranges (see Section~\ref{section-UphillPrediction}).
\end{enumerate}

These results highlight that the approximate solution provided by the P-SHDL model, where the corrective flux $J^C_i$s are estimated with the analytical equilibrium expression \eqref{eqn:equilibrium_mpnp_scaled1}, is quite accurate in predicting the potential, charge, and flux profiles. Self-consistency of the electrostatics plays a major role in shaping the solution of these Poisson drift diffusion models with volume exclusion phenomena.

\begin{figure}[ht!]
    \centering
    \includegraphics[width=\linewidth]{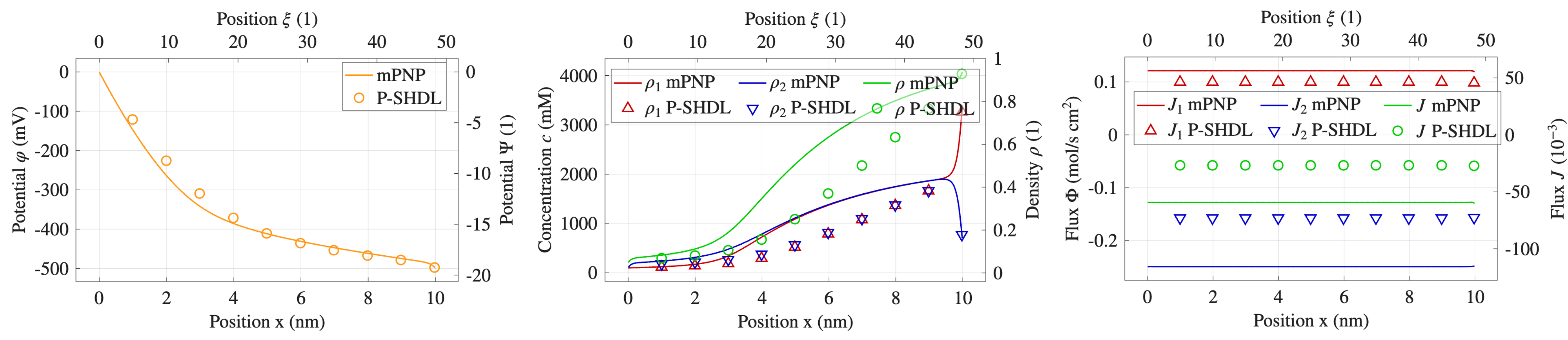}
    \caption{Potential, concentration and fluxes profiles computed with the mPNP and P-SHDL models when $\rho^R\simeq 1$ and the difference between the boundary densities of each species is large. In this setting, the two profiles display a noticeable mismatch, showing that the equilibrium approximation is not valid anymore. Boundary conditions: $\Psi^R = 19.34$, $\Psi^L = 0$, $\rho_1^R=0.7867$, $\rho_2^R=0.1349$, $\rho_1^L=\rho_2^L=0.022$, corresponding to $\varphi^R = 500\,$mV, $\varphi^L = 0\,$mV, $c_1^R=3500\,$mM, $c_2^R=600\,$mM, $c_1^L=c_2^L=100\,$mM, respectively.}
    \label{fig:break}
\end{figure}

However, as seen in Figure \ref{fig:break}, the agreement between the mPNP and P-SHDL models breaks down when the right boundary conditions are even closer to saturation than in Figure\,\ref{fig:HV-HC}; namely, when $\rho_{1}^{R}+\rho_{2}^{R}\simeq 1$, and when the difference between boundary values $|\rho_{i}^{R}-\rho_{i}^{L}|$, $i=1,2$, increases. Indeed, under these circumstances, the system is too far from the equilibrium assumption that yields the same expression for the corrective flux 
of P-SHDL and mPNP models
(cf. \eqref{eq:SumOfRhoDerivatives}).

\section{Uphill transport for charged particle models}\label{Sect:Uphill-Section}

In this Section, we focus on the onset of uphill transport regimes in the context of the SHDL, mPNP, P-SHDL, and PNP models previously introduced. In particular, we discuss the roles played by steric (volume) exclusion, the Poisson equation, and the equilibrium assumption.

Firstly, we look at the appearance of uphill transport in conditions as for the phase diagrams of Section \ref{Sect:UphillForSHDL} and for the results in Section \ref{Section-comparison-OutOfEq}, see Figure \ref{fig:comparisonMath}. 
Then, we analyze the physical case study, depicted in Fig. \ref{fig:membrane_sketch}, which represents an embodiment of the general system in Fig. \ref{fig:basic_sketch} whose essential physics can be modeled with the SDHL, mPNP and P-SHDL. 

\subsection{Uphill regimes in the SHDL, mPNP, P-SHDL and PNP models}
\label{section-UphillPrediction}

We have seen in Section \ref{Section-comparison-OutOfEq} that the non-self-consistent, non-equilibrium solution provided by the SHDL model is still useful since the signs of the partial fluxes coincide with the ones of the mPNP model. 

Here, we expand the analysis comparing the uphill phase diagram for the SHDL model reported in Figure~\ref{fig:fase 1-NOSAT} with the predictions of the P-SHDL and mPNP models, where the Poisson equation is included, under identical density boundary conditions and for a domain of unitary normalized length. Given \eqref{potential-assumptions} (used to reduce the SHDL transport equations to the P-SHDL ones), this comparison spans the anti-diagonal of the phase diagrams shown in Section \ref{sect:UphillPhaseDiagrams} (i.e., $a = -b$), where the right-boundary potential $\Psi(\xi^{R})\in[-2, 2]$. 

\begin{figure}[ht!]
\begin{subfigure}{0.3\textwidth}
\includegraphics[width=\linewidth]{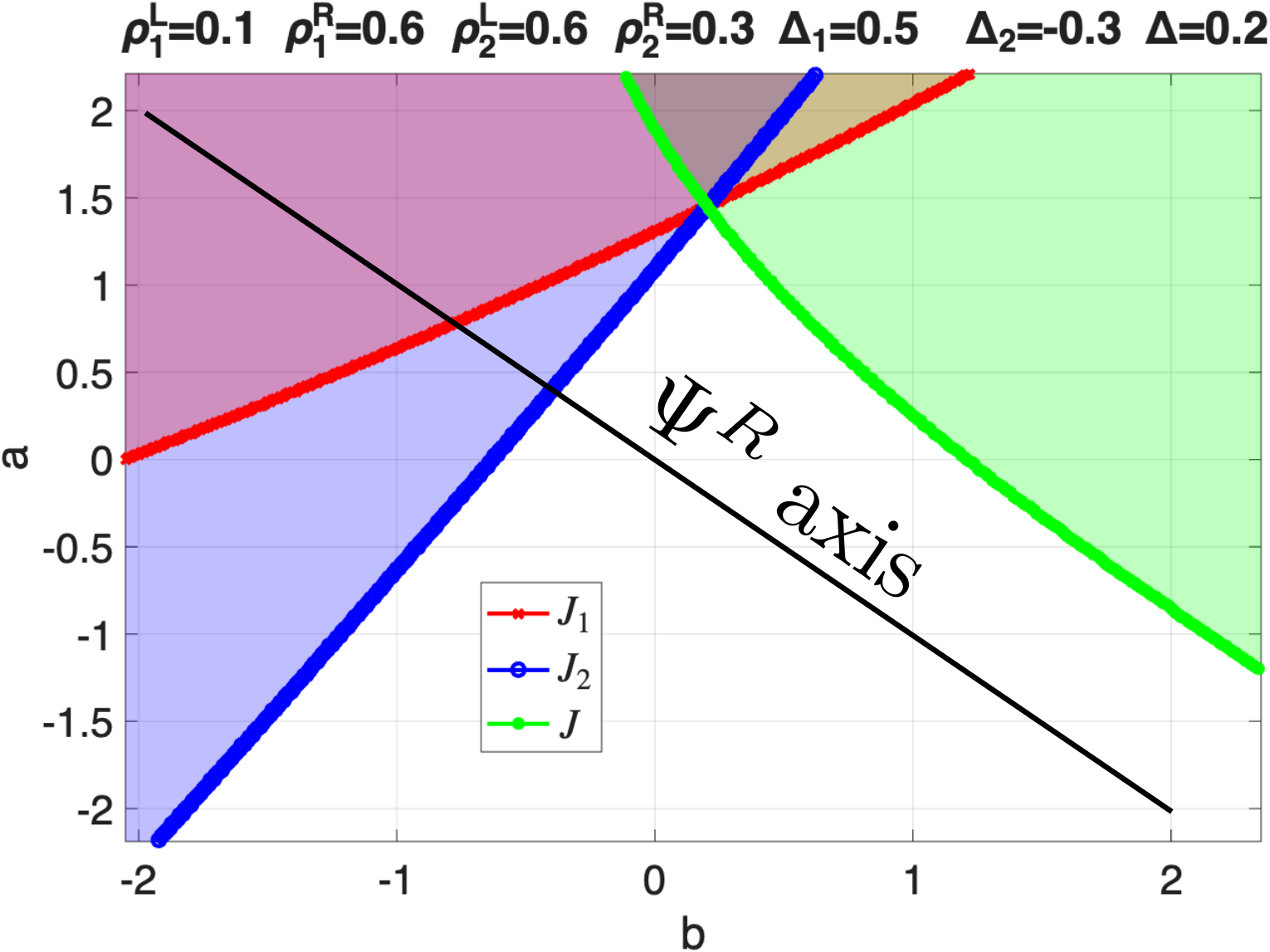}
	\centering
    \caption{Uphill diagram for the SHDL model (same as in Figure \ref{fig:fase 1-NOSAT}). }
	\label{fig:fase 2-NOSAT_II}
\end{subfigure}
\hspace{0.03\textwidth}
\begin{subfigure}{0.3\textwidth}
  \centering
    \includegraphics[width=1\linewidth]{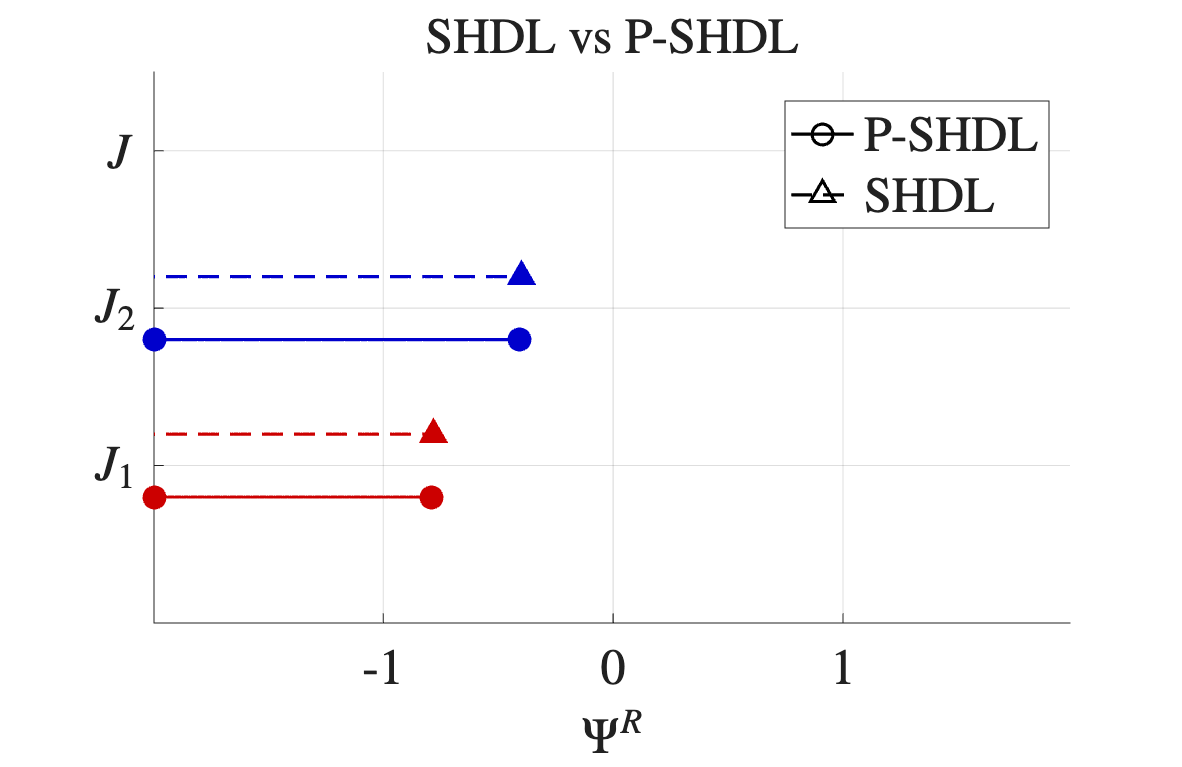}
    \caption{Parameter ranges where uphill transport is present according to the P-SHDL and SHDL models.}
    \label{fig:UpredP-SHDLMath}
\end{subfigure}
\hspace{0.03\textwidth}
\begin{subfigure}{0.3\textwidth}
  \centering
    \includegraphics[width=1\linewidth]{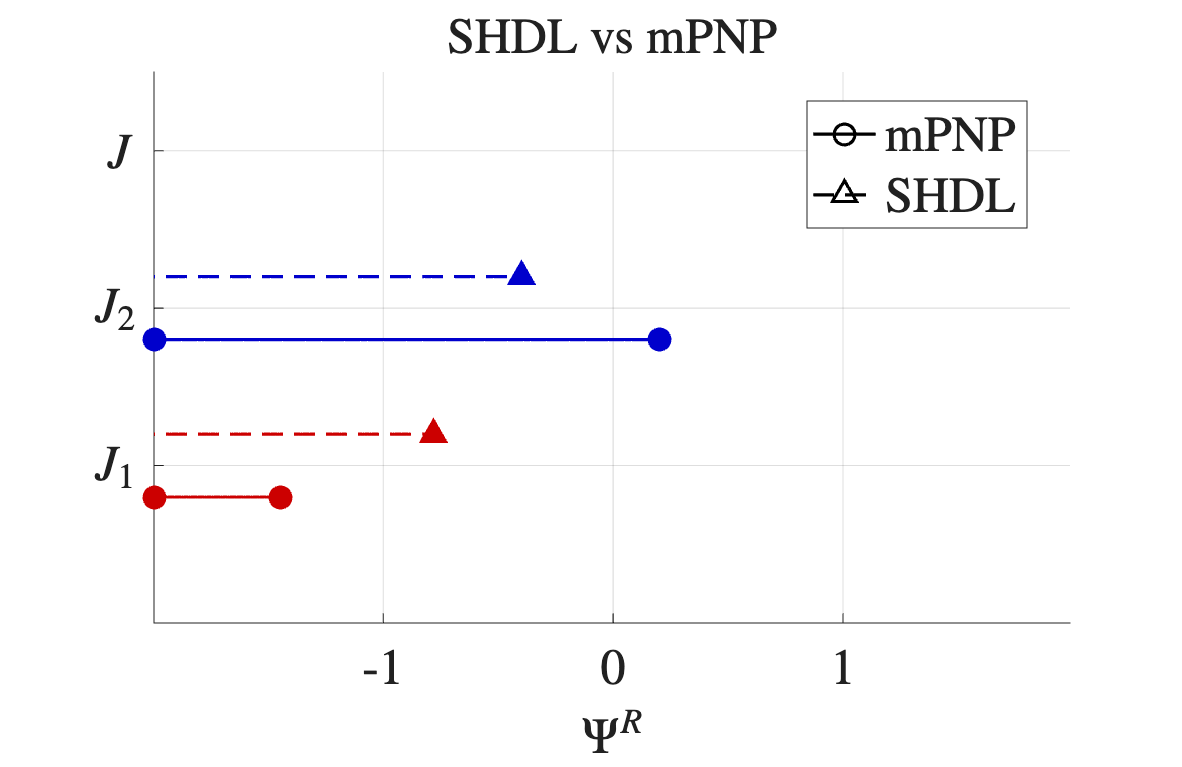}
    \caption{Parameter ranges where uphill transport is present according to the mPNP and SHDL models.}
        \label{fig:UpredmPNPMath}
    \end{subfigure}
    \caption{Left: Uphill transport phase diagram with the SHDL model as in Figure \ref{fig:fase 1-NOSAT}. The black diagonal scans the parameter line $b=-a$ on which direct comparison with the P-SHDL and mPNP models is carried out. Center and right: Parameter ranges where uphill transport exists according to the P-SHDL model (center) and the mPNP model (right). Same conditions as in Fig. \ref{fig:fase 1-NOSAT}. ; $\xi^{R}=1$, $\Psi(\xi^{R})\in$$[-2, 2]$; hence,  $a = -b\in$$[-2, 2]$, $\rho_{1}^{L}=0.1$, $\rho_{1}^{R}=0.6$, $\rho_{2}^{L}=0.6$ and $\rho_{2}^{R}=0.3$ as in Figure~\ref{fig:fase 2-NOSAT_II}. 
}
\label{fig:comparisonMath}
\end{figure}

Figure~\ref{fig:comparisonMath} demonstrates a good quantitative agreement between the $\Psi^R$ ranges where uphill transport takes place according to the SHDL and the P-SHDL model (panel (b)), while the correspondence with the mPNP model is comparatively weaker (panel (c)).

This result is not surprising. The chosen simulation parameters, $\rho^L$=0.7 and $\rho^R$=0.9, correspond to strong exclusion phenomena. Furthermore, the concentration gradients are very large, but the total space charge density is small due to the small length of the domain. As a result, the RHS of the Poisson equation is nearly zero and solving the Poisson equation is not so important because the field is anyhow almost constant. This explains why the P-SHDL and the SHDL yield similar predictions in Figure \ref{fig:UpredP-SHDLMath}. However, the system is very far from equilibrium due to large concentration gradients and fields. Consequently, the mPNP model, which does not assume equilibrium, departs from the P-SHDL (and SHDL, Figure \ref{fig:UpredmPNPMath}) models.

\subsection{Physical case study: ion permeable membrane}\label{Sect: CaseStudy1}

\begin{figure}[ht!]
    \centering
    \includegraphics[width=0.5\linewidth]{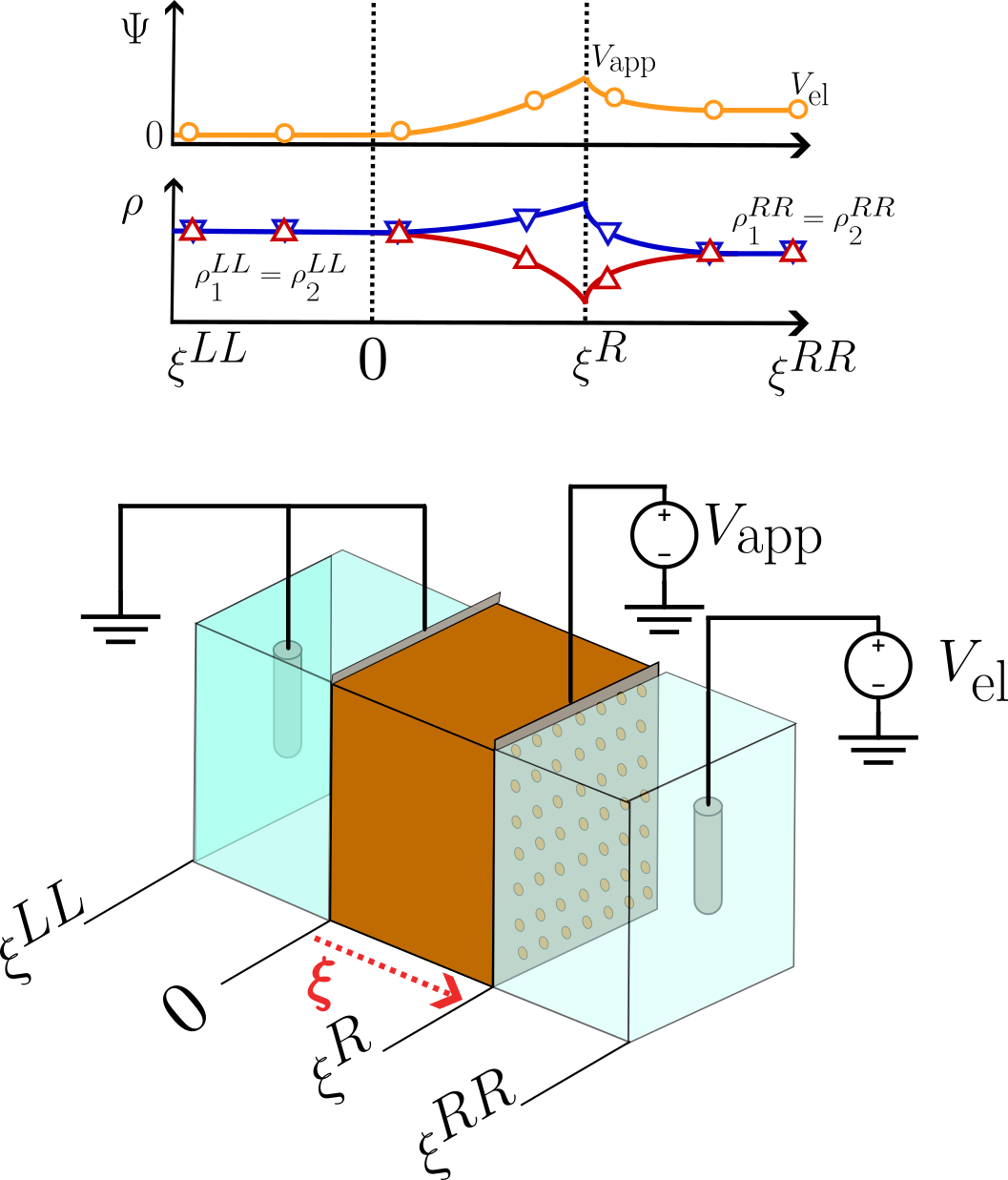}
    \caption{Sketch of the system for the physical case study: an ion-permeable membrane is contacted by two thin, porous, and ideally polarizable electrodes in contact with two electrolytes, each with its own reference electrode.}
    \label{fig:membrane_sketch}
\end{figure}

Figure \ref{fig:membrane_sketch} represents an ion-permeable membrane (in brown) between two bulk regions of symmetric electrolytes biased at constant voltages by suitable power supplies and reference electrodes. The potential at the electrolyte/membrane interface is set by thin, porous, and ideally polarizable electrodes, allowing ionic flow across them. 

The electrolyte region on the left is grounded to maintain electroneutrality; i.e., a constant $c_1^{LL}=c_2^{LL}$ which sets the left boundary condition $\rho^L$ for the membrane domain. To focus on the essentials of the physical problem, the membrane space charge density is neglected, and the partition factor to the nearby electrolytes is set to $1$.
The membrane is biased at the voltage $V_{app}$ by the ideally polarizable electrodes.
The voltage drop $(V_{app}-V_{el})$, applied across the electrolyte region on the right, unbalances the ions' concentration from the charge neutrality value $c_1^{RR}=c_2^{RR}$ at the location of the reference electrode to the values $c_1^{R}\neq c_2^{R}$ at the interface with the membrane. Therefore, by properly choosing the electrolytes' salinity, $c_1^{LL}=c_2^{LL}$ and $c_1^{RR}=c_2^{RR}$ and voltages, $(V_{app}$ and $V_{el})$, it is then possible to set the concentrations and potentials at the membrane boundaries, with the only constraint that $c_1^R$ and $c_2^R$ are not independent, since they must obey equations \eqref{eqn:equilibrium_mpnp1} and \eqref{eqn:equilibrium_mpnp2} for the same $\varphi^R=(V_{app}-V_{el})$. 

The membrane domain resembles the one in Fig. \ref{fig:basic_sketch} and can be described by the mPNP, P-SHDL or SHDL equations, with the electrolyte regions imposing the boundary conditions. 

\begin{figure}[ht!]
    \centering
    \includegraphics[width=\linewidth]{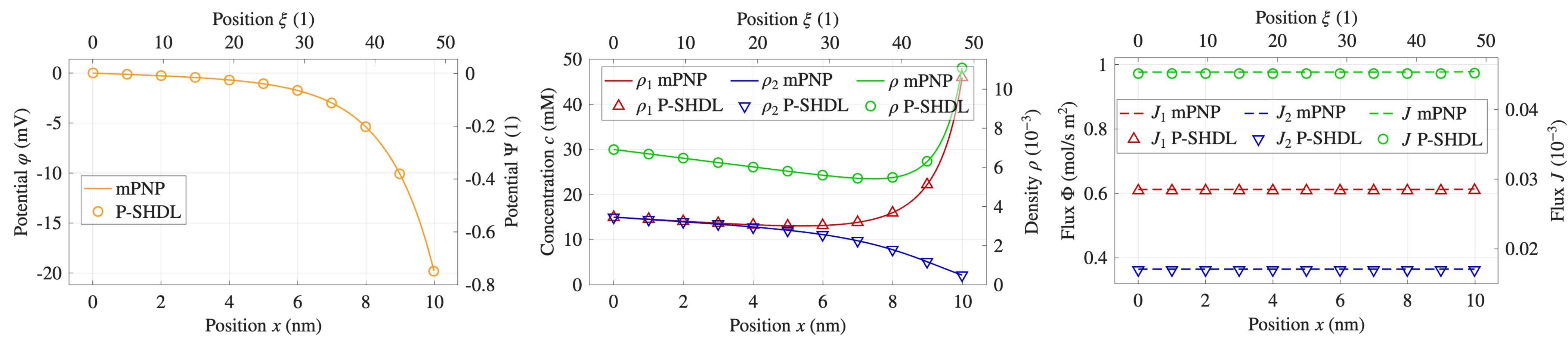}
    \caption{Potential, concentrations, and fluxes in the domain of the membrane computed with the P-SHDL and mPNP models. $c_1^{LL} = c_2^{LL} = c^{LL} = 15\,$mM and $c_1^{RR} = c_2^{RR} = 10\,$mM. $V_{app}=-20 \,$mV; $V_{el}=0$\,V. $\xi^{R}=10\,$nm. The concentrations at $\xi^{R}$ are computed according to equations \eqref{eqn:equilibrium_mpnp1} and \eqref{eqn:equilibrium_mpnp2}.}
    \label{fig:membrane_plots}
\end{figure}

Figure \ref{fig:membrane_plots} shows the results of mPNP and P-SHDL simulations at low voltage and low concentration, where self-consistency with the Poisson equation, exclusion phenomena, and saturation have modest impacts. The space dependence of the relevant physical quantities is similar to those observed in Figures \ref{fig:HV-HC} and \ref{fig:LV-LC}. The fluxes are constant, as expected. The excellent agreement between mPNP and P-SHDL models is consistent with previous results. Comparing the sign of the concentration gradients and the fluxes, we find global uphill transport and partial uphill for species 1 above $\xi\simeq 30$, and no uphill for species 2. These results suggest a deeper investigation. 

\begin{figure}
    \centering
    \includegraphics[width=\linewidth]{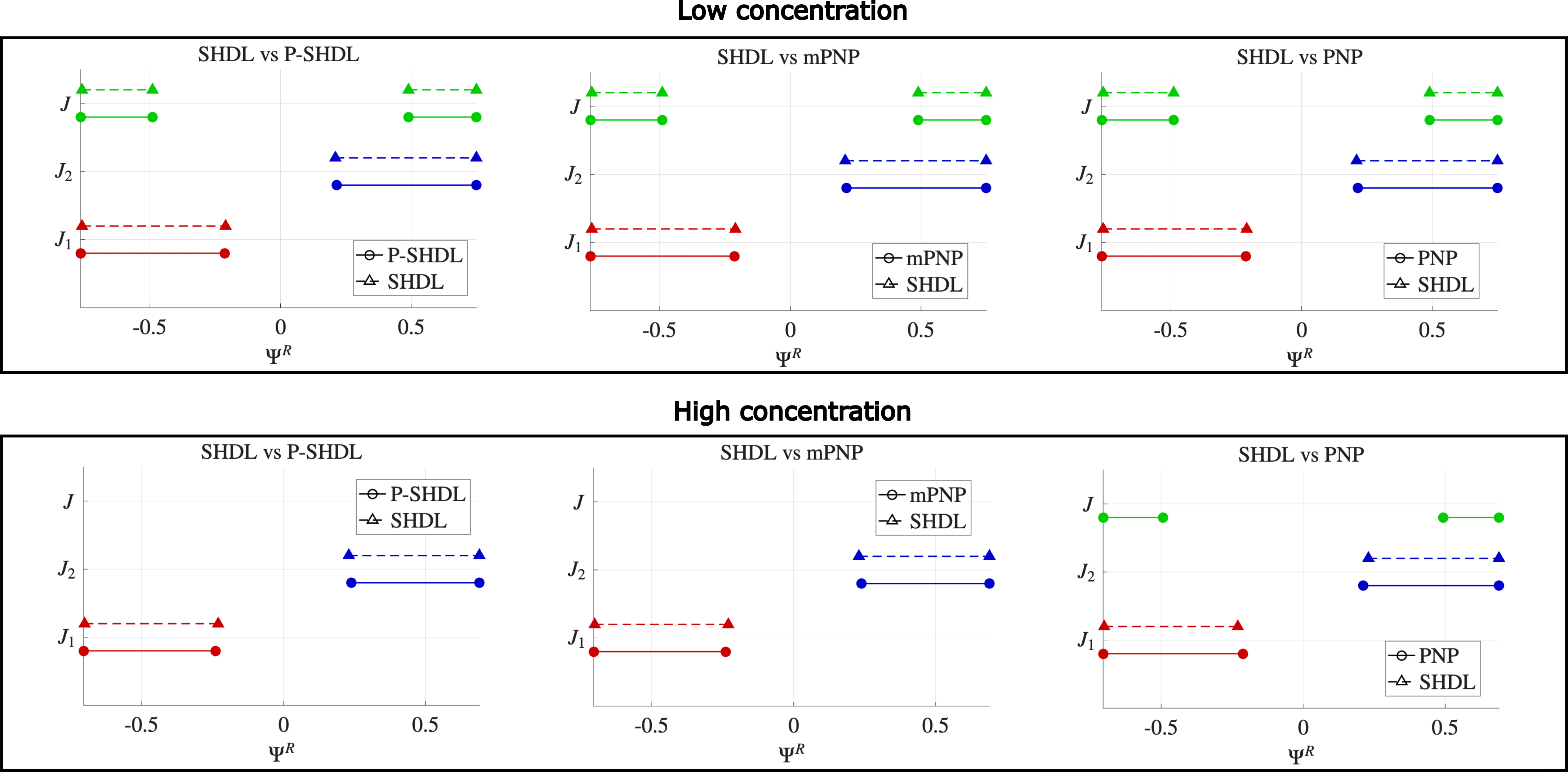}
    \caption{$\Psi^R$   where uphill transport exists in  the system in Figure \ref{fig:membrane_sketch} according to the P-SHDL, mPNP, SHDL, and PNP models. Top panel: low salinity (diluted electrolyte, $c^{LL}$=15 mM, $c^{RR}$=10 mM, same as Fig.\,\ref{fig:membrane_plots}). Bottom panel: high salinity (concentrated electrolyte, $c^{LL}$=1500 mM, $c^{RR}$=1000 mM). The PNP model is the only one that does not account for steric effects. The SHDL model is the only one that does not solve the Poisson equation. }
    \label{fig:antidiagonal}
\end{figure}

To this end, the segments in Figure \ref{fig:antidiagonal} show the $\Psi^R$ ranges where partial and global uphill transport is present according to the different models. The top panels refer to low electrolytes' salinity with the same concentrations as in Figure \ref{fig:membrane_plots} (i.e., diluted system); the bottom panels are for high concentrations and electrolytes' salinity. For the SHDL case, $\Psi^R$ spans the antidiagonal of the uphill maps as done in figure \ref{fig:fase 2-NOSAT_II}. We observe, as expected, that species 1 and 2 exhibit uphill transport in opposite ranges of $\Psi^R$ values. 

At low concentrations and fluxes (top panels), all models (including the common PNP where exclusion is neglected in the domain and at the boundary) are in excellent agreement in predicting the $\Psi^R$ ranges where uphill transport exists. This comes as no surprise, because at low concentrations the potential is essentially linear; hence, the coincidence of results for the P-SHDL and SHDL models. Moreover, quasi-equilibrium holds, which yields coincidence between the mPNP and P-SHDL models; finally, exclusion phenomena are negligible, hence, the agreement between the mPNP and the common PNP model. 
If, instead, the concentrations increase by, e.g., two orders of magnitude (bottom panel of Figure \ref{fig:antidiagonal}), the remarkable impact of exclusion phenomena causes the failure of the PNP model (compare the bottom center and bottom right graphs), also because without exclusion the boundary values of the $\rho_i$ are quite different for the PNP from those of the mPNP. 

The remarkable difference in global uphill region predictions of simulations at low and high concentration can be better understood by scrutinizing the diffusive ($J^F$), drift ($J^D$), and corrective ($J^C$) contributions to the total flux. 

\begin{figure}[ht!]
\begin{subfigure}{0.5\textwidth}
\centering
	\includegraphics[width=\linewidth]{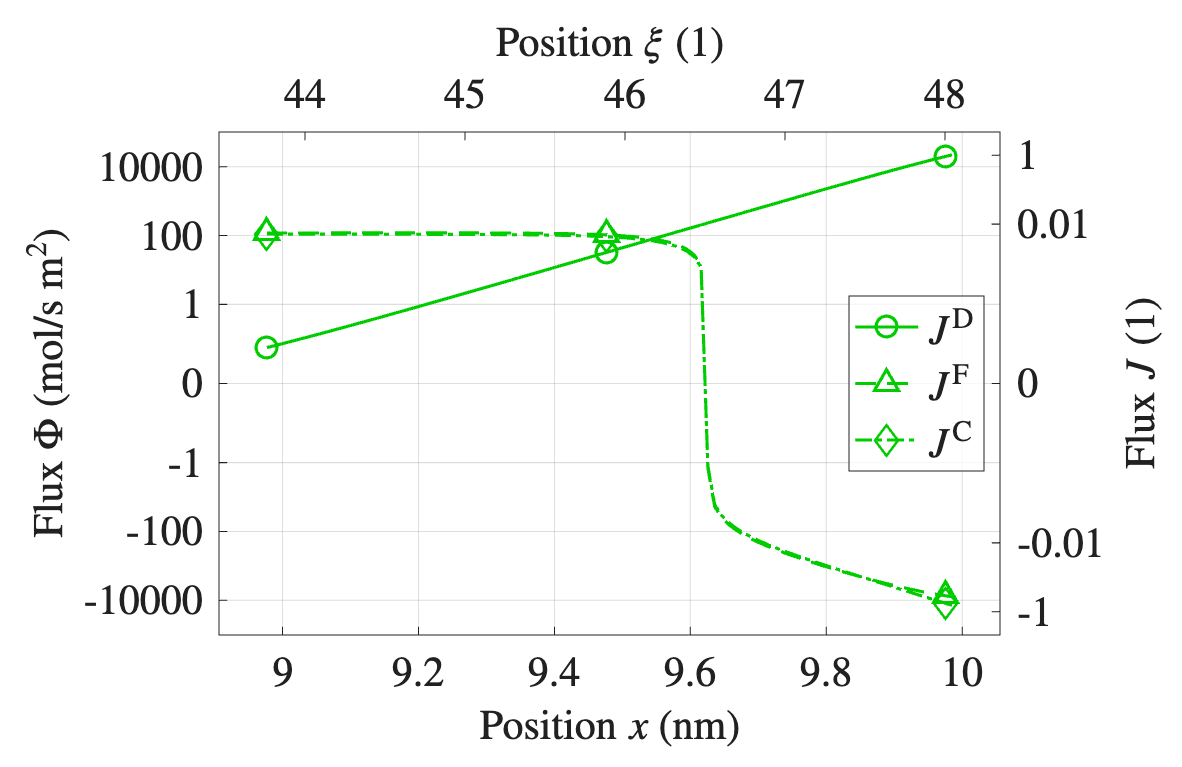}
	\caption{High concentrations \label{fig:high_conc_sep}}
\end{subfigure}
\begin{subfigure}{0.5\textwidth}
\centering
	\includegraphics[width=\linewidth]{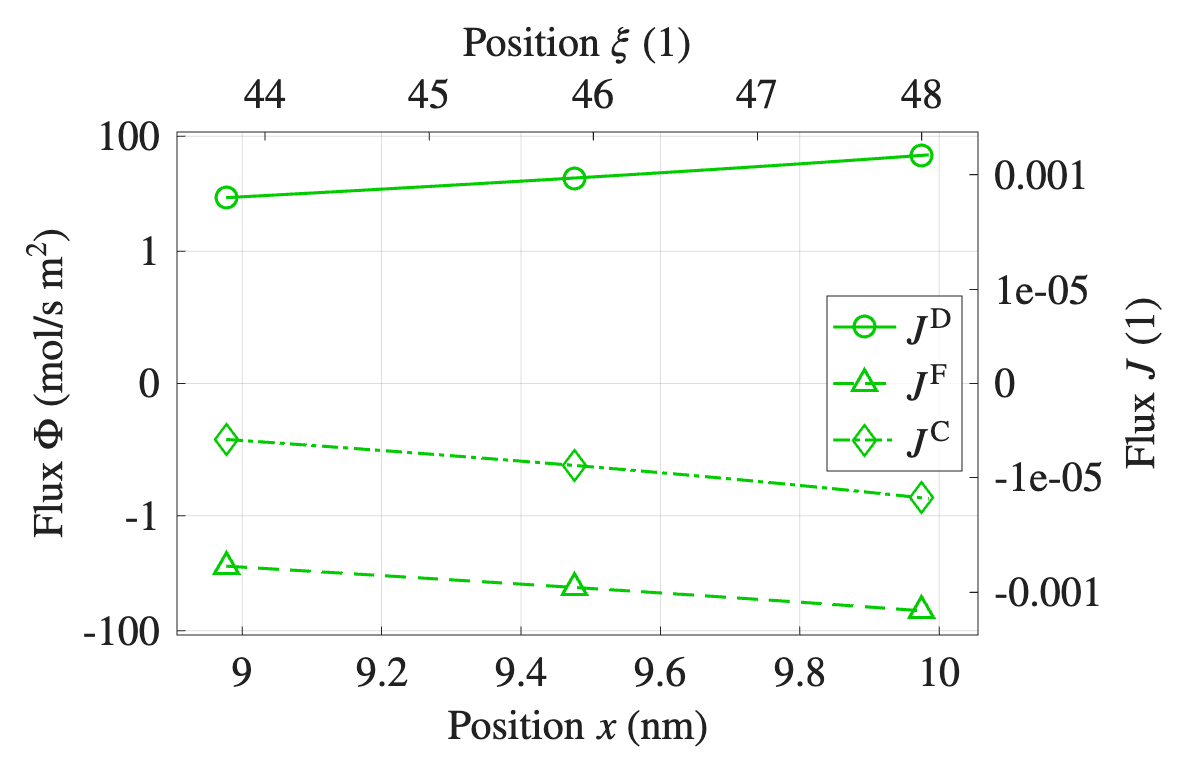}
	\caption{Low concentrations \label{fig:low_conc_sep}}
\end{subfigure}
   \caption{Contribution of the Fick, drift, and steric exclusion corrective fluxes to the total global ion flux for the mPNP model in arcsinh scales. The left and right plots refer to the high boundary concentration and low boundary concentration cases in Fig.\,\ref{fig:antidiagonal}, respectively. $\varphi^R=$-20\,mV, i.e. $\Psi^R$=-0.77 corresponding to the leftmost boundary of all the plots in \ref{fig:antidiagonal}.}  
   \label{fig:separated_fluxes}
\end{figure}

Fig. \ref{fig:separated_fluxes} shows the profiles of the global flux components next to the right boundary of the domain, computed with the mPNP model for the low and high concentration cases in Fig. \ref{fig:antidiagonal} and $\varphi^R=-20$\,mV, i.e. $\Psi^R$=-0.77. In addition, Table\,\ref{Tab:separatefluxes} reports the magnitude of all flux components and the concentration boundary driving $\Delta$ at $\xi^R$. 

If we focus on the total global fluxes (Table \ref{Tab:separatefluxes}), we see that at low concentrations (Fig. \ref{fig:low_conc_sep}) the $J^C$s are smaller in absolute value than all other components by about two orders of magnitude. At high concentrations (Fig. \ref{fig:high_conc_sep}), the magnitude of the correction term is comparable with the other components. When the concentration of the electrolyte rises, the global total flux maintains the same sign while the boundary driving $\Delta$ changes sign, thus causing the absence of global uphill transport. 
In the high concentration case, the saturation of the rightmost electrolyte causes the global concentration to be lower than at the leftmost electrolyte, so $\Delta$ becomes negative. Instead, in the PNP case, concentrations have no upper bound, so $\rho^R$ can grow to a level higher than $\rho^L$ and ensure $\Delta>0$ and uphill transport. 

If we now observe the partial fluxes' magnitude in Table \ref{Tab:separatefluxes}, we see that at low concentration, and for both species, the $J^C_i$s are smaller in absolute value than any other component and are negative, while $J^F_i$ and $J^D_i$ have opposite signs. In particular, $J^C_2$ has the same sign as $J^D_2$, thus favoring the creation of partial uphill transport conditions for species 2, while the opposite is found for species 1.   
At high concentration, instead, the corrective terms related to exclusion phenomena are of the same order or even larger than the Fick ones but the $|J^D_i|$ is still very relevant in determining the presence of partial uphill. 

These observations lead to the conclusion that, for charged species, self-consistency between the electrostatics is a mandatory ingredient, without which the density and potential profiles are highly inaccurate; although, if the concentration gradients of all species are imposed, then the qualitative presence of uphill transport may still be correctly predicted.  
In all cases of charge transport, neglecting exclusion at high concentrations can completely miss the presence or absence of partial or global uphill, as well as the actual intensities of the fluxes. A detailed study of all these situations would be very interesting; however, it goes beyond the scope of the present contribution and will be the subject of a future work.

\begin{table}
\centering
\caption{Partial and global unitless flux components at $\xi^R$ for the simulations in Figure\,\ref{fig:separated_fluxes}.}
\label{Tab:separatefluxes}
\renewcommand{\arraystretch}{1.3}
\begin{tabular}{ccccccc}
\hline
\textbf{Low conc.}&
\textbf{$\Delta = \rho^R-\rho^L$}&
\textbf{$J^F$}&
\textbf{$J^D$}&
\textbf{$J^C$}&
\textbf{$J$}&
\textbf{Uphill}
\\
\hline
\textbf{species 1}&
$0.007$&
$-2.267 \cdot 10^{-3}$&
$2.318 \cdot 10^{-3}$&
$-2.234\cdot 10^{-5}$&
$2.850\cdot 10^{-5}$&
Yes
\\
\hline
\textbf{species 2}&
$-0.003$&
$1.267\cdot 10^{-4}$&
$-1.087\cdot 10^{-4}$&
$-1.048\cdot 10^{-6}$&
$1.693\cdot 10^{-5}$&
No
\\
\hline
\textbf{global}&
$0.004$&
$-2.140 \cdot 10^{-3}$&
$2.209\cdot 10^{-3}$&
$-2.339\cdot 10^{-5}$&
$4.543\cdot 10^{-5}$&
Yes

\\
\hline
\textbf{High conc.}&
\textbf{$\Delta = \rho^R-\rho^L$}&
\textbf{$J^F$}&
\textbf{$J^D$}&
\textbf{$J^C$}&
\textbf{$J$}&
\textbf{Uphill}
\\
\hline
\textbf{species 1}&
$0.266$&
$-0.476$&
$1.098$&
$-0.617$&
$0.005$&
Yes
\\
\hline
\textbf{species 2}&
$-0.301$&
$0.108$&
$-0.066$&
$-0.037$&
$0.005$&
No
\\
\hline
\textbf{global}&
$-0.035$&
$-0.368$&
$1.032$&
$-0.654$&
$0.010$&
No
\\\end{tabular}
\end{table}

\section{Conclusions}\label{Sect:Conclusions}

In this work, we have examined uphill transport in systems where drift–diffusion dynamics are strongly shaped by finite-size exclusion, a regime particularly relevant for nanoscale electrochemical environments. 
Starting from the hydrodynamic limit of a multispecies weakly asymmetric exclusion process (HDL), we have shown that the resulting stationary model (SHDL) naturally predicts regimes in which particle fluxes invert their direction and flow \textit{according} to the imposed density gradient, and not the \textit{opposite}. These inversion regimes arise from the nonlinear interplay of diffusion with the field-driven drift and steric pressure fluxes, and admit a clear characterization in terms of phase diagrams in the space of external drivings. 

We then focus on systems of charged particles and identify well-defined hypotheses under which the SHDL model, further endowed with the Poisson equation (P-SHDL), converges to the modified Poisson-Nernst-Planck model adopted in several fields of physics. In particular, we have shown that the mPNP and the P-SHDL are equivalent sets of equations if the corrective terms generated by volume exclusion are estimated using charge density expressions valid at equilibrium. This approximation of the mPNP clarifies the validity limits of the P-SHDL and entails the possibility to express the corrective terms as a function of the density gradients (Equations \eqref{mPNP}), instead of the fields (Equation \eqref{eqn:exclusion-currents}).  

By comparing the SHDL predictions with those obtained from the modified Poisson–Nernst–Planck (mPNP) equations, the self-consistent SHDL model (P-SHDL), and the bare PNP model where exclusion is neglected, we have demonstrated that the essential mechanisms behind exclusion-driven uphill transport are preserved across models. In particular, the agreement between SHDL and P-SHDL uphill predictions is remarkably strong. 
At the same time, the general mPNP model reveals how non-equilibrium drift and exclusion reshape flux inversion, especially in dense or highly saturated regimes.

The analysis of a representative template nanoscale system involving charged particles, where the electric field and average concentration gradients can be set separately (an ion-selective membrane between two electrolytes), illustrates how electrostatic interactions and exclusion processes control the onset, amplitude, and suppression of uphill transport in practical settings. In both cases, crowding modulates global density, alters local electric fields, and can either favor or reduce partial or global flux inversion and uphill transport when saturation is approached.

Overall, our results clarify the microscopic origin, mathematical structure, and range where there is physical relevance of uphill transport in crowded environments, and bridge the gap between exclusion-based particle models and continuum electrochemical descriptions. They also suggest that uphill transport may play a significant role in nanoscale electrolytes, confined ionic and iontronic devices, and membrane-based technologies.
\subsection*{Acknowledgment}
 \thanks{This work received partial support by PNRR M4 C2 INV 1.5, NextGenerationEU, Avviso 3277/2021, ECS\textbackslash\_00000033 ECOSISTER spoke 6, and by the EU- NextGenerationEU, Italian MUR under PNRR- M4 C2-I1.3 Project PE 00000019 ``HEAL ITALIA'' CUP E93C22001860006, spoke 1. JN and LS acknowledge useful discussions with Prof. P.Palestri, Universit\`a di Modena e Reggio Emilia. CG acknowledges financial support from INDAM through INdAM Project 2024 CUP E53C23001740001.}
\bibliography{reference}
\bibliographystyle{unsrt}
\newpage
\appendix
{\huge \textbf{Appendix}}
\section{Uphill transport with zero boundary driving}\label{appendix-A}
In this appendix, we report simulation results for the special setup in which the difference of the global boundary densities is zero as a consequence of $\Delta_{1}=-\Delta_2$. In this case, we may define uphill transport as every setup exhibiting a non-zero global current. 

\begin{figure}[h!]
\begin{subfigure}{0.5\textwidth}
\includegraphics[width=0.8\linewidth]{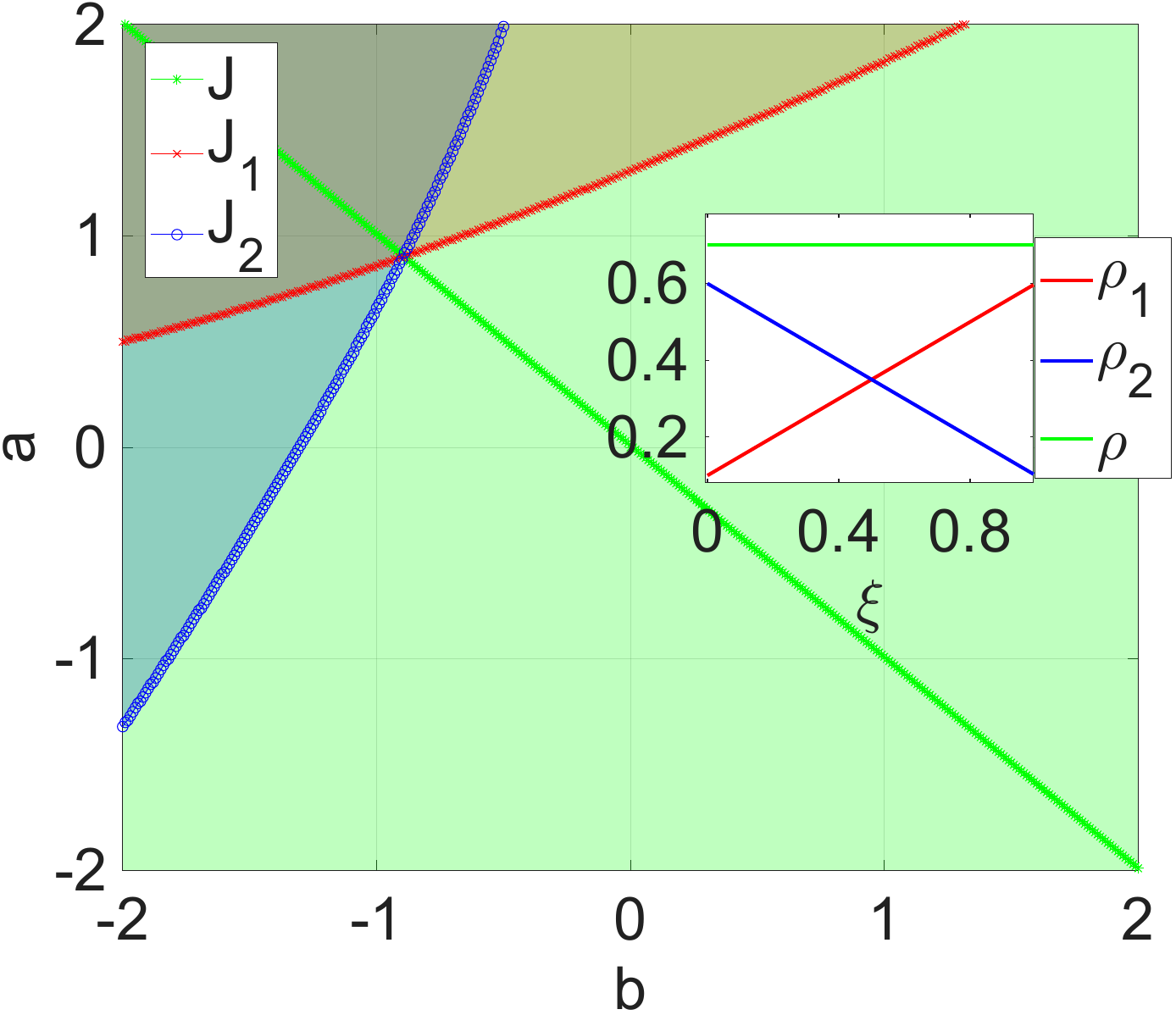}
	\centering\captionsetup{width=0.8\linewidth}
    \caption{Case without saturation at the boundaries. $\rho_{1}^{L}=0.1$, $\rho_{1}^{R}=0.6$, $\rho_{2}^{L}=0.6$, $\rho_{2}^{R}=0.1$, $\Delta_{1}=0.5$, $\Delta_{2}=-0.5$, $\Delta=0$.}
	\label{fig:fase 2-NOSATDegenere}
\end{subfigure}
\begin{subfigure}{0.5\textwidth}
\centering\captionsetup{width=0.8\linewidth}
	\includegraphics[width=0.8\linewidth]{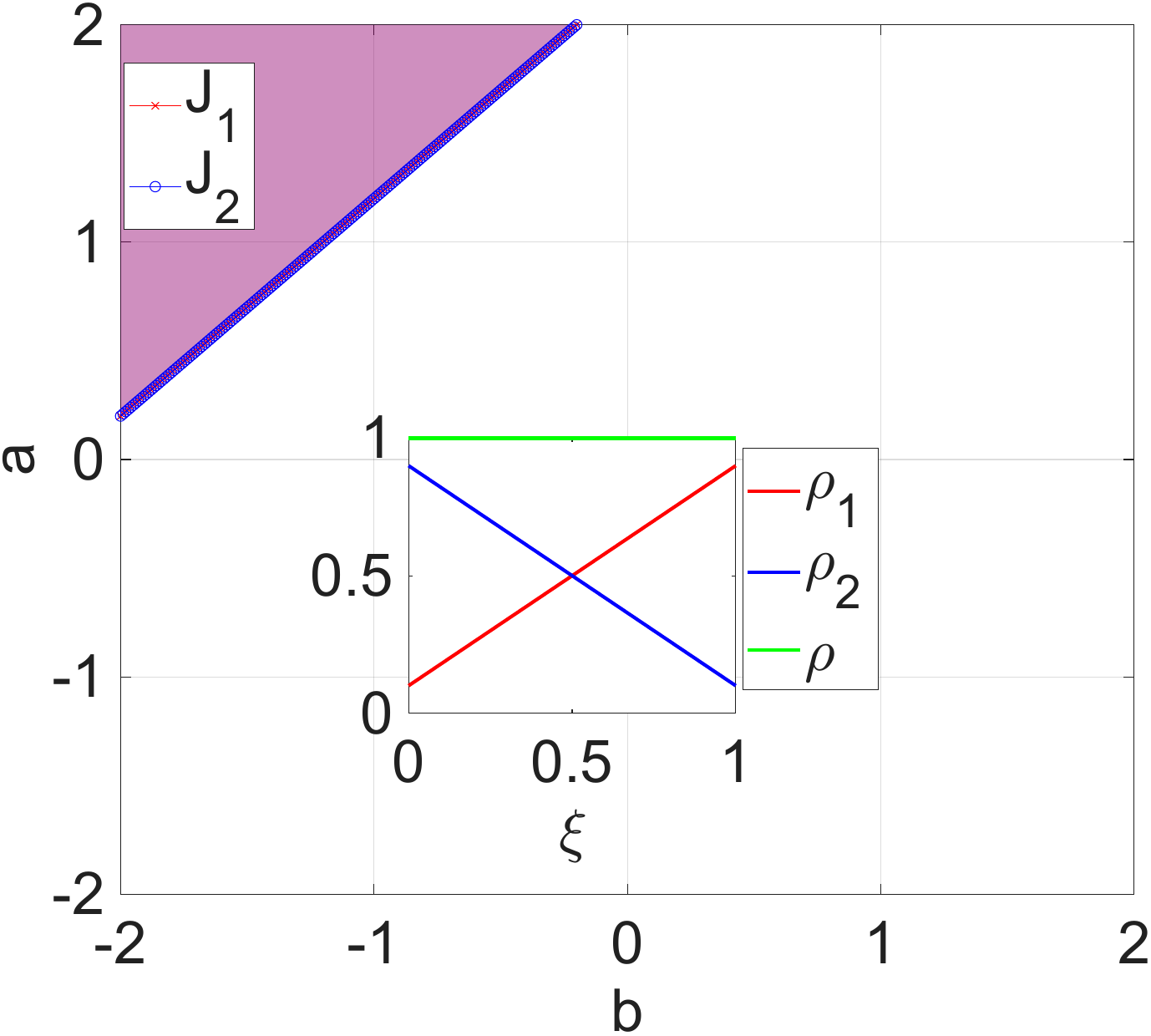}
	\caption{Case with saturation at the right boundary. $\rho_{1}^{L}=0.1$, $\rho_{1}^{R}=0.9$, $\rho_{2}^{L}=0.9$, $\rho_{2}^{R}=0.1$,    $\Delta_{1}=0.8$, $\Delta_{2}=-0.8$, $\Delta=0$}
	\label{fig:fase 2-SATDegenere}
\end{subfigure}
   \caption{
The plots represent the lines of sign change (symbols), and the regions of uphill transport (shaded areas) for the flux of species~1 (red), species~2 (blue), and for the global flux (green), in the space of parameters $a,b$ (external fields). The insets show the qualitative trend of the partial and global densities. }
\label{fig:image2Degenere}
\end{figure}

\paragraph{Simulation set \#3: $\Delta=0$, $\Delta_{1}>0$, $\Delta_{2}<0$ and $\Delta_{1}=-\Delta_{2}$.}
Here, we look for the following types of uphill transport:
\begin{itemize}
	\item \textit{Partial uphill for the species 1:} since $\Delta_{1}>0$, we have uphill transport if $J_{1}>0$;
	\item \textit{Partial uphill for the species 2:} since $\Delta_{2}<0$, we have uphill transport if $J_{2}<0$;
	\item \textit{Global uphill:}
	since $\Delta=0$, we have uphill transport if 	$J\neq 0$.

\end{itemize}

Figure \ref{fig:fase 2-NOSATDegenere} shows the results of the numerical simulations in the form of phase diagrams for a set of boundary densities such that $\rho^{L}=\rho^{R}<1$. In this case, we identify regions of the graph where more than one uphill regime (partial or global) coexist. Furthermore, global uphill (green shaded area) is present everywhere up to the anti-diagonal of the plane, that is, except along the green line where the global flux is zero. 

Figure \ref{fig:fase 2-SATDegenere} shows the results of the numerical simulations for the set of boundary conditions in which $\rho^{L}=\rho^{R}=1$. Here, the phase diagram shows only two regions: a region where both the partial uphill for species $1$ and $2$ are possible; a region where no uphill is present. No global uphill is visible due to a vanishing global mobility (saturation).

\end{document}